\documentclass[twocolumn,english]{revtex4-1}
\usepackage[T1]{fontenc}
\usepackage[latin9]{inputenc}
\usepackage{color}
\usepackage{units}
\usepackage{amsmath}
\usepackage{amssymb}
\usepackage{graphicx}
\usepackage{esint}
\usepackage{caption}
\makeatletter

\providecommand{\tabularnewline}{\\}

\usepackage{braket}

\makeatother

\usepackage{babel}
\begin{document}

\title{\textcolor{black}{Environmental design principles for efficient excitation
energy transfer in dimer and trimer pigment-protein molecular aggregates
and the relation to non-Markovianity}}

\author{Charlotta Bengtson}

\affiliation{Department of Chemistry - \AA ngstr\"om Laboratory, Theoretical Chemistry,
Uppsala University, Box 538, Uppsala, SE-751 21, Sweden.}

\author{Michael Sahlin}

\affiliation{Department of Physics, Stockholm University, AlbaNova University
Center, SE-106 91 Stockholm, Sweden.}

\author{and Marie Ericsson}

\affiliation{Department of Physics and Astronomy, Materials Theory, Uppsala University,
Box 516, Uppsala, SE-751 20, Sweden.}

\date{1 January 2018}
\begin{abstract}
Lately there has been an interest in studying the effects and mechanisms
of environment-assisted quantum transport, especially in the context
of excitation energy transfer (EET) in pigment-protein molecular aggregates.
\textcolor{black}{Since these systems can be seen as open quantum
systems where the dynamics is within the non-Markovian regime, the
effect of non-Markovianity on efficient EET as well as its role in
preserving quantum coherence and correlations has also been investigated
in recent works. In this study, we explore optimal environments for
efficient EET between end sites in a number of dimer and trimer model
pigment-protein molecular aggregates when the EET dynamics is modeled
by the HEOM-method. For these optimal environmental parameters, we
further quantify the non-Markovianity by the BLP-measure to elucidate
its possible connection to efficient EET. We also quantify coherence
in the pigment systems by means of the measure $l_{1}-$norm of coherence
to analyze its interplay with environmental effects when EET efficiency
is maximal. Our aim is to investigate possible environmental design
principles for achieving efficient EET in model pigment-protein molecular
aggregates and to determine whether non-Markovianity is a possible
underlying resource in such systems. We find that the structure of
the system Hamiltonian (i.e., the pigment Hamiltonian parameter space)
and especially, the relationship between the site excitation energies,
determines whether one of two specific environmental regimes is the
most beneficial in promoting efficient EET in these model systems.
In the first regime, optimal environmental conditions are such that
the EET dynamics in the system is left as coherent as possible. In
the second regime, the most advantageous role of the environment is
to drive the system towards equilibrium as fast as possible. In reality,
optimal environmental conditions may involve a combination of these
two effects. We cannot establish a relation between efficient EET
and non-Markovianity, i.e., non-Markovianity cannot be regarded as
a resource in the systems investigated in this study.}

\noindent \ 

\end{abstract}
\maketitle

\section{Introduction}

The time evolution of a quantum system interacting with an environment
is of interest in many research fields studying open quantum systems,
such as quantum information theory and condensed matter physics \citep{Breuer}.
Generally, the interaction with a macroscopic environment, i.e., an
environment consisting of infinitely many degrees of freedom, has
mainly been seen as a source of dissipation and decoherence in the
system, leading to destruction of desirable quantum resources such
as quantum entanglement. The main focus of technological developments
of quantum systems for applications such as quantum computation has
consequently been to isolate the system from its environment. 

Recently, it has been recognized that in some quantum systems, environmental
effects may help to protect, or even create, quantum resources in
form of quantum correlations and coherence in the system. Well known
examples of such systems are certain photosynthetic complexes, each
consisting of a network of pigments (light absorbing organic molecules)
held together by a protein scaffold, forming a molecular aggregate.
Despite the fact that the network of pigments is strongly coupled
to a macroscopic environment at physiological temperature, long-lasting
quantum coherence - which may act as a resource in these complexes
- has been experimentally verified \citep{Engel2007,Hayes2010,Panitchayangkoon2010,Panitchayangkoon2011}. 

In a photosynthetic complex, one of the pigments absorbs a photon
which is transferred as excitation energy through the network of pigments
until it reaches a pigment that is functioning as an end site. In
connection to this pigment in the network, the excitation is captured
by a reaction center and converted to chemical energy. The overall
efficiency of photosynthetic complexes hence depends, among other
things, on how efficient an excitation can be transferred from the
initially excited pigment to the pigment in contact with the reaction
center. Photosynthetic complexes are known to convert excitation energy
to chemical energy with high efficiency \citep{Chain1977} which has
created an interest in studying their properties and trying to mimic
their features in artificial photosynthesis-technology. It is believed
that quantum coherence together with environmental effects enables
such an efficient excitation energy transfer (EET) among the pigments
in the network and many studies have tried to reveal how quantum coherence,
different environmental interactions and EET efficiency are related
to each other \citep{Qin2014,Plenio2008,Caruso2009,Chin2010,Sinayskiy2012,Marais2013,Wu2010,Rebentrost2009_1,Rebentrost2009_2,Mohseni2008,Dijkstra2012,Shabani2012,Mohseni2014}. 

The standard computational treatment to study the effects of an environment
on an open quantum system, which has been used in the study of pigment-protein
molecular aggregates in Refs. \textcolor{black}{\citep{Rebentrost2009_2,Plenio2008,Caruso2009,Wu2010,Rebentrost2009_1,Mohseni2008}},
has been to employ Markovian master equations to model the dynamics.
The Markovian approximation is to assume that the environment is in
equilibrium during the time evolution of the system. For some processes,
such as EET in photosynthetic complexes where the strength of the
system-environment interaction is of the same order as the intra-system
interactions, Markovian master equations are not sufficient to capture
relevant environmental effects. In order to accurately take those
effects into account, the dynamics has to be modeled by master equations
within the non-Markovian regime. As a result, there is a possibility
for information to flow back-and-forth between the system and the
environment during the time evolution of the system. Hence, environmental-induced
revivals of quantities such as coherence and correlations within the
system may occur. 

In the case of photosynthetic complexes, the need for non-Markovian
master equations to capture the behavior of the system-environment
interactions in a realistic manner has resulted in developments of
new theoretical methods. Especially the hierarchical equations of
motion (HEOM) method, originally derived in Refs. \citep{Tanimura1989_1,Tanimura1989_2,Tanimura1990}
and then extended in Refs. \citep{Ishizaki2009_2,Ishizaki2009_1},
has served as the benchmarking method to model EET in photosynthetic
complexes.

Non-Markovian dynamics has been found to be a resource in some processes.
Examples include entanglement generation \citep{Paz2008,Valido2013_1,Valido2013_2,Bellomo2007,Huelga2012},
information transfer in a noisy channel \citep{Maniscalco2007,Bylicka2014,Caruso2014}
and quantum communication \citep{Laine2014,Liu2013}. In Ref. \citep{Thorwart2009}
it is shown that non-Markovianity can function as a resource for prolonging
the duration of coherence in a dimer pigment-protein model system.
Non-Markovian effects in pigment-protein molecular aggregates have
also been studied in Refs. \citep{Chen2014,Rebentrost2011,Mujica-Martinez2013}. 

As in the case of any potential quantum resource, it is crucial to
be able to quantify non-Markovian effects in a system in order to
be able to distinguish different environments in terms of their ability
to function as a resource. Indeed, a number of different measures
of non-Markovianity has been developed and proposed \citep{Breuer2009,Rivas2010,Rajogopal2010}.
One that has been frequently used in studies of non-Markovianity is
the BLP-measure \citep{Breuer2009}. It is based on the observation
that the distinguishability of two initial states can never increase
under a Markovian evolution. The distinguishability is quantified
by the trace distance and an increase in the trace distance in a certain
time interval is interpreted as a backflow of information from the
environment to the system. 

In this study, we investigate numerically how the parameters describing
the environment and its coupling to the system influence EET efficiency
between end sites in a network of $N$ ($N=2,3$) pigments in a protein
scaffold when the HEOM method is used to model the EET dynamics. Within
a range of possible environmental parameters, we seek for an optimum
in the efficiency of the EET. We investigate a number of pigment configurations
(defined by the pigments excitation energies and inter-pigment couplings)
in order to evaluate how optimal environmental parameters may differ
depending on the system under consideration. We further determine
whether the dynamics imposed by the optimal environmental parameters
(with respect to EET efficiency) is within the non-Markovian regime
or not. To explore the interplay between coherence and environmental
effects, as well as the possible connection between coherence and
non-Markovianity, we quantify the amount of coherence in our systems
by the $l_{1}$-norm of coherence \citep{Baumgratz2014}.

This study \textcolor{black}{attempts to reveal design principles
on how to create environmental conditions for efficient EET in pigment-protein
molecular aggregates. Further, we aim to evaluate the possible role
of non-Markovianity for efficient EET in such aggregates. The results
can hopefully be useful for gaining insights on how to design artificial
molecular aggregates for light harvesting. }

\section{Systems}

\begin{figure}
\includegraphics[scale=0.2]{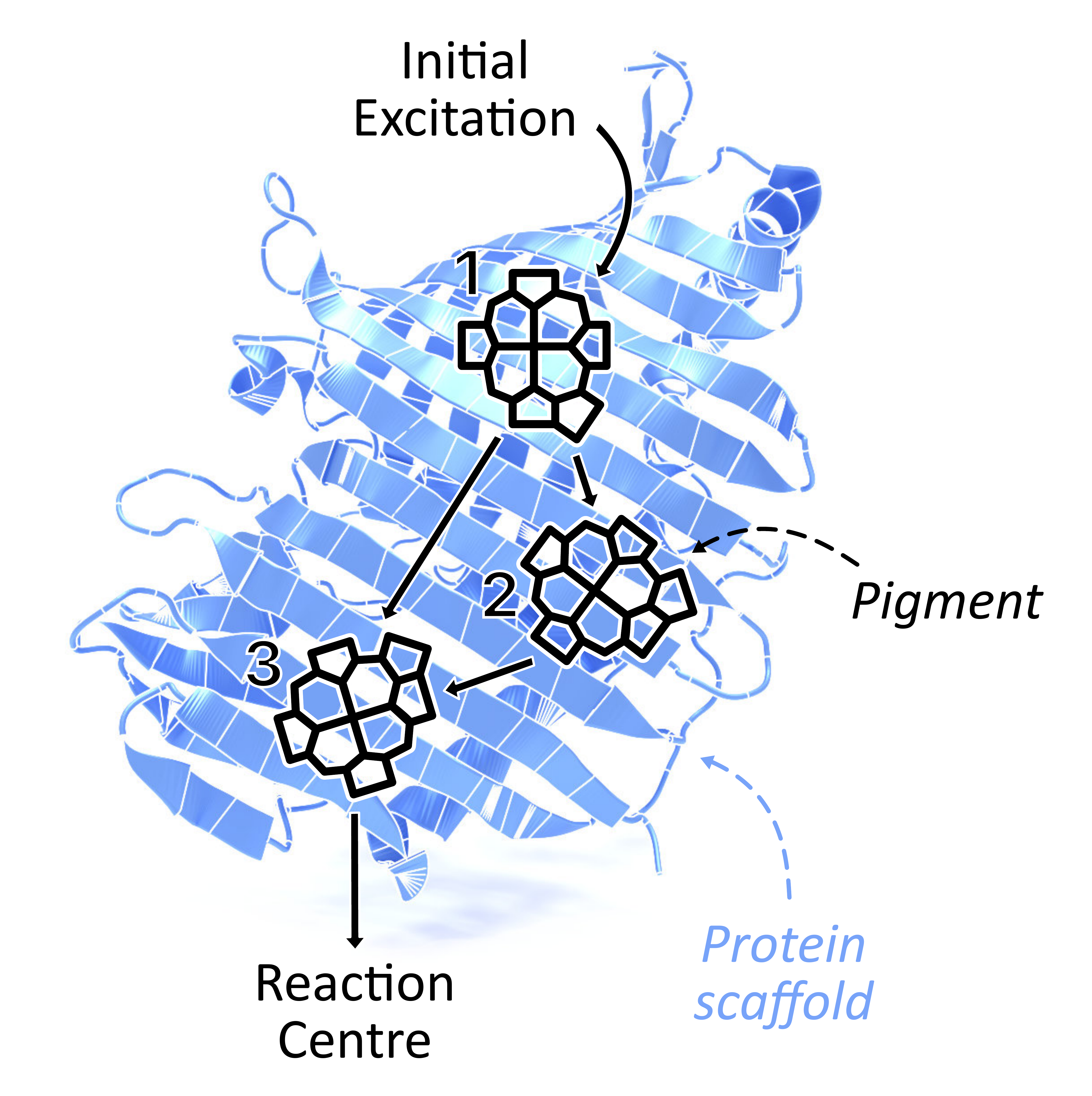}\protect\caption{\label{fig:Trimer}Schematic illustration of \textcolor{black}{EET
in} a trimer pigment-protein molecular aggregate.}

\end{figure}

The systems considered in this study are networks of ${\color{red}{\color{black}N}}$
($N=2,3$) sites where each site represents a pigment. These two types
of systems can capture both pure quantum tunneling between pigments
(dimer, i.e., $N=2$) and interference between different pathways
(trimer, i.e., $N=3$), which are two important mechanisms in EET.
A general trimer system ($N=3$) is illustrated in Fig. \ref{fig:Trimer}
along with the EET direction and possible EET pathways. The illustration
also shows how the pigments are attached to a protein scaffold, which
acts as the environment. 

Each pigment is modeled as a two-level-system consisting of a ground
state, $\ket{\Psi_{g}}$, and an excited state, $\ket{\Psi_{e}}$.
The Hamiltonian of the full $N$-site system is assumed to be given
by \citep{Renger2001} 

\begin{equation}
\hat{H}_{S}=\underset{i}{\sum}E_{i}\ket{i}\bra{i}+\underset{i\neq j}{\sum}J_{ij}\ket{i}\bra{j},\label{eq:Hamiltonian}
\end{equation}

\noindent where $E_{i}$ is the excitation energy of site $i$, i.e.,
the energy required to excite site $i$ from $\ket{\Psi_{g}}$ to
$\ket{\Psi_{e}}$, and $J_{ij}$ is the coupling (in reality, the
electrostatic dipole-dipole coupling) of the two sites. The state
$\ket{i}$ is equivalent to the state $\ket{\Psi_{g}^{1}\ldots\Psi_{e}^{i}\ldots\Psi_{g}^{N}}$,
i.e., pigment $i$ is in its excited state while the rest of the pigments
in the aggregate are in their ground states.

A well known natural pigment-protein molecular aggregate is one of
the monomers in the Fenna-Matthews-Olson (FMO) complex \citep{Fenna1974}\textcolor{black}{.
It} is commonly used as a model system for photosynthetic EET \citep{Caruso2010,Sarovar2010,Fassioli2010,Baker2015,Chen2013,Rebentrost2009_2,Plenio2008,Caruso2009,Chin2010,Marais2013,Wu2010,Rebentrost2009_1,Mohseni2008,Dijkstra2012,Shabani2012,Mohseni2014,Shabani2014,Wu2012,Mujica-Martinez2013,Rebentrost2011}
and consists of seven pigments arranged in such a way that there are
two different routes for an initial excitation to be transferred to
the reaction centre.\textcolor{black}{{} The first of these two routes
consists of pigments $1$, $2$ and $3$, where the initial excitation
is located on pigment $1$ and pigment $3$ is in contact with the
reaction centre, and the second route involves pigment $4$, $5$,
$6$ and $7$ \citep{Wu2012}. Because the s}econd route has an energy
downhill structure, while the first route has an energy barrier between
the first two sites which might require quantum tunneling, we choose
to look at a model system consisting of $N$ sites that captures the
features of the first EET route in a FMO-complex monomer. For the
dimer case, these are: $\left|E_{1}-E_{2}\right|\sim\left|J_{12}\right|$.
We refer to this system Hamiltonian as $\hat{H}_{{\rm {FMO}}}^{(2)}$.
In the trimer case, we further have $E_{1,2}>E_{3},$ $\left|J_{12}\right|>\left|J_{23}\right|>\left|J_{13}\right|$,
with Hamiltonian referred to as $\hat{H}_{{\rm {FMO}}}^{(3)}.$ 

In a recent study \citep{Bengtson2017}, the optimal Hamiltonian parameter
space for a dimer and trimer pigment aggregate with respect to EET
efficiency and time-averaged coherence in a closed system were found.
We also investigate these systems, whose Hamiltonians are referred
to as $\hat{H}_{{\rm {E}}}^{(N)}$ and $\hat{H}_{{\rm {C}}}^{(N)}$,
respectively, where $N=2,3$.\textcolor{red}{{} }\textcolor{black}{The
specific Hamiltonian parameters for all six systems investigated in
this study can be found in Tab. \ref{tab:Dimer} (dimers) and Tab.
\ref{tab:Trimer} (trimers).}

\noindent 
\begin{table}
\protect\caption{\label{tab:Dimer}Hamiltonian parameters for dimer systems. \textcolor{black}{All values are given in units of ${\rm {cm}}^{-1}$.}}
\begin{tabular}{|c|c|c|}
\hline 
\textbf{\textcolor{black}{Hamiltonian}} & $\mathbf{E_{1}-E_{2}}$ & $\mathbf{J_{12}}$\tabularnewline
\hline 
\hline 
$\hat{H}_{{\rm {FMO}}}^{(2)}$ & $-100$ & $-100$\tabularnewline
\hline 
$\hat{H}_{{\rm {E}}}^{(2)}$ & $0$ & $100$\tabularnewline
\hline 
$\hat{H}_{{\rm {C}}}^{(2)}$ & $144$ & $100$\tabularnewline
\hline 
\end{tabular}
\end{table}

\noindent 
\begin{table}
\protect\caption{\label{tab:Trimer}Hamiltonian parameters for trimer systems. \textcolor{black}{All
values are given in units of ${\rm {cm}}^{-1}$.}}
\begin{tabular}{|c|c|c|c|c|c|}
\hline 
\textbf{\textcolor{black}{Hamiltonian}} & $\mathbf{E_{1}-E_{3}}$ & $\mathbf{E_{2}-E_{3}}$ & $\mathbf{J_{12}}$ & $\mathbf{J_{23}}$ & $\mathbf{J_{13}}$\tabularnewline
\hline 
\hline 
$\hat{H}_{{\rm {FMO}}}^{(3)}$ & $200$ & $300$ & $-100$ & $50$ & $0$\tabularnewline
\hline 
$\hat{H}_{{\rm {E}}}^{(3)}$ & $0$ & $0$ & $100$ & $100$ & $0$\tabularnewline
\hline 
$\hat{H}_{{\rm {C}}}^{(3)}$ & $40$ & $-160$ & $-100$ & $-20$ & $-100$\tabularnewline
\hline 
\end{tabular}
\end{table}

\section{Modeling and optimizing the EET efficiency}

\noindent 
\begin{figure}
\includegraphics[scale=0.4]{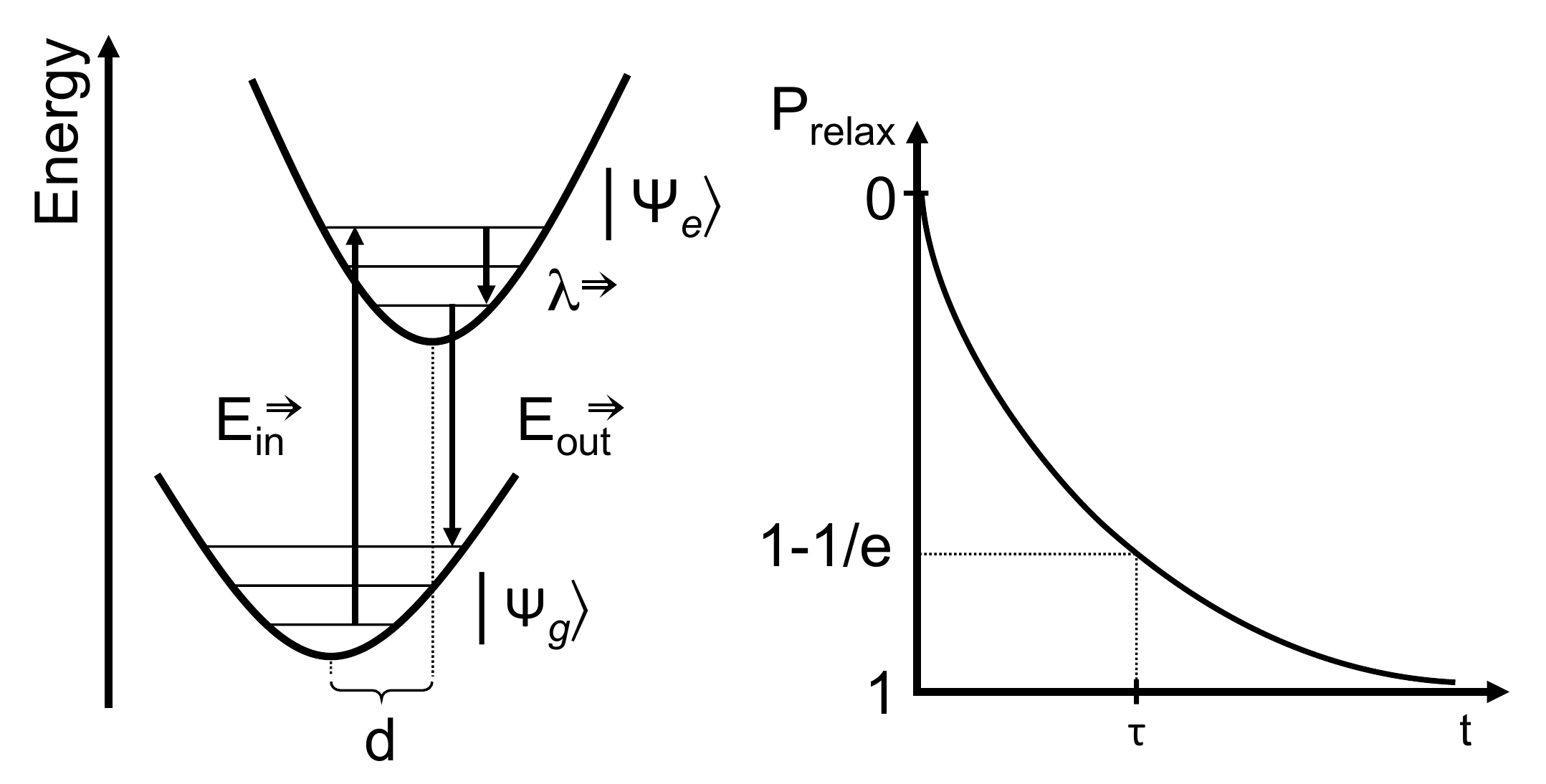}\protect\caption{\label{fig:Lambda_gamma}Physical interpretation of the parameters
in HEOM. Left: Excitation from the ground state, $\ket{\Psi_{g}}$,
to the excited state, $\ket{\Psi_{e}}$, of a pigment. Since the transition
occurs vertically, the excitation of the electronic state is accompanied
by excitation of vibrational (phonon) states. Relaxation of the vibrationally
excited states releases reorganization energy, $\lambda$, which is
proportional to the displacement, $d$, of the equilibrium configuration.
Right: The parameter $\tau$ governs the decay rate of vibrationally
excited states. At $t=\tau$, there is a probability ($P_{{\rm {relax}}}$
in the figure) of $1-\nicefrac{1}{e}$ that all vibrationally excited
states have relaxed to their vibrational ground states.}
\end{figure}

The EET in the systems is modeled using the HEOM-method \citep{Ishizaki2009_2,Ishizaki2009_1}
where the parameters representing the environment and its coupling
to the system are varied in order to optimize EET efficiency. In the
derivation of the HEOM method, the protein environment is modeled
as a set of harmonic oscillator modes, i.e., phonons. EET between
sites in the network occurs via non-equilibrium phonon states in accordance
with vertical Franck-Condon transitions. The phonons are locally and
linearly coupled to each site and when they relax to their equilibrium
states, \textcolor{black}{reorganization energy}\textit{\textcolor{black}{,}}
denoted by $\lambda$, is released to the environment. Excitation-
and de-excitation processes, including the release of the reorganization
energy, of a pigment are shown in Fig. \ref{fig:Lambda_gamma}. The
reorganization energy characterizes the strength of the system-environment
coupling; a low value of $\lambda$ corresponds to a coherent EET
dynamics. 

The phonon relaxation dynamics is characterized by the parameter $\tau=\gamma^{-1}$,
which determines the time scale of the environmental fluctuations.
The physical meaning of $\tau$ is shown in Fig. \ref{fig:Lambda_gamma},
where it can be seen that the smaller $\tau$ is, the faster will
a major part of the phonons relax to equilibrium. The environment
is further described by its temperature, $T$, which defines the thermal
equilibrium of the environment. \textcolor{black}{Even though each
site has its own, local environment, the parameters describing the
environment ($\lambda,$ $\tau$ and $T$) are taken to be the same
for all sites. A summary of the environmental parameters used in HEOM
are found in Tab. \ref{tab:Environmental-parameters}. }

\textcolor{black}{The density operator, $\hat{\rho}$, describing
the EET dynamics in the network of pigments is found by solving a
set of HEOM, where the number of coupled equations are determined
by a truncation condition \citep{Ishizaki2009_2,Ishizaki2009_1}.
This condition neglects phonon states much higher in energy than the
characteristic frequency of the pigment-system. }

\textcolor{black}{The ranges of the temperature and the reorganization
energy are chosen to be}

\textcolor{black}{
\begin{equation}
250\,{\rm {K}}\leq T\leq300\,{\rm {K}}
\end{equation}
and}

\textcolor{black}{
\begin{equation}
20\,{\rm {cm}^{-1}}\leq\lambda\leq220\,{\rm {cm}^{-1}}.
\end{equation}
In this way we interpolate between system-environment coupling strengths
corresponding to EET dynamics in the coherent regime (where $\lambda$
is significantly smaller than the strongest inter-site couplings)
and those corresponding to the incoherent regime (where $\lambda$
is significantly larger than the strongest inter-site couplings).
The temperature is chosen to be within natural conditions and below
temperatures where the protein scaffold would denature and hence,
the 3D-structure of the pigment-protein aggregate would be destroyed.
The range of the parameter $\tau$ is chosen such that the high temperature
condition is fulfilled, which is a requirement to use HEOM in the
form presented here \citep{Ishizaki2009_2,Ishizaki2009_1}, i.e.,}

\textcolor{black}{
\begin{equation}
\tau>\hbar\beta,\:\beta=k_{B}T,\label{eq:High T limit}
\end{equation}
where $k_{B}$ is Boltzmann's constant. At $T=250\,{\rm {K}}$ (which
is the critical temperature according to Eq. \ref{eq:High T limit}
in this case), the above condition requires that $\tau>31\,{\rm {fs}}.$
To be within this regime with clear margin, we use}

\textcolor{black}{
\begin{equation}
50\,{\rm {fs}}\leq\tau\leq500\,{\rm {fs}}
\end{equation}
in our calculations. Due to numerical limitations, the parameter step-sizes
are $\Delta T=2.5\,{\rm {K}},\:\Delta\lambda=10\,{\rm {cm}^{-1}},\:\Delta\tau=25\,{\rm {fs}}$
in the dimer systems and $\Delta T=5\,{\rm {K}},\:\Delta\lambda=20\,{\rm {cm}^{-1}},\:\Delta\tau=50\,{\rm {fs}}$
in the trimer systems. Further, we consider EET on the time scale
of $1\:{\rm {ps}}$, i.e., $0\:{\rm {fs}}\leq t\leq1000\:{\rm {fs}},$
which is a time scale widely accepted for EET in photosynthetic complexes. }

\textcolor{black}{Excitation of site $1$ is taken as initial condition
for the EET dynamics. The convergence of the numerical solution is
tested at critical values of $\lambda,$ $\tau$ and $T$ prior to
performing the calculations for the full parameter ranges. }

\textcolor{black}{We also compare the time evolution of the site-populations
to the corresponding populations at equilibrium ($t\rightarrow\infty$),
which is defined by the system Hamiltonian (without environment),}

\textcolor{black}{
\begin{equation}
\hat{\rho}_{{\rm {eq}}}=\frac{e^{-\beta\hat{H}_{S}}}{Z},\label{eq:Equilibrium}
\end{equation}
}

\textcolor{black}{
\begin{equation}
Z={\rm {Tr}}(e^{-\beta\hat{H}_{S}}).
\end{equation}
}

\noindent 
\begin{table}
\protect\caption{\label{tab:Environmental-parameters}Summary of environmental parameters
used in HEOM.}
\begin{tabular}{|c|c|c|}
\hline 
\textbf{\textcolor{black}{Parameter description}} & \textbf{\textcolor{black}{Symbol}} & \textbf{\textcolor{black}{Unit}}\tabularnewline
\hline 
\hline 
System-environment coupling & $\lambda$ & ${\rm {cm}^{-1}}$\tabularnewline
\hline 
Environmental timescale & $\tau$ & ${\rm {\rm {fs}}}$\tabularnewline
\hline 
Environmental temperature & $T$ & ${\rm {K}}$\tabularnewline
\hline 
\end{tabular}

\end{table}

\section{Quantifying efficiency, non-Markovianity and coherence}

In this section we describe how we quantify EET efficiency (section
\ref{sub:Efficiency}), non-Markovianity (section \ref{sub:Non-Markovianity})
and coherence (section \ref{sub:Coherence}).

\subsection{\label{sub:Efficiency}Quantifying efficiency}

\noindent The efficiency for transferring the initial state, $\hat{\rho}_{i}=\ket{1}\bra{1}$,
into the target state, $\hat{\rho}_{f}=\ket{N}\bra{N}$, is quantified
by \textit{fidelity, $F$,}\textit{\textcolor{black}{{} }}\textcolor{black}{\citep{Nielsen}}
which takes the form

\begin{equation}
F(\hat{\rho})={\rm {Tr}}\sqrt{\sqrt{\hat{\rho}_{f}}\hat{\rho}\sqrt{\hat{\rho}_{f}}},\label{eq:Fidelitet_mixade tillst=0000E5nd}
\end{equation}
where $\hat{\rho}$ is the resulting state when $\hat{\rho}_{i}$
is evolved in time for a certain set of parameters $T$, $\lambda$
and $\tau$. Since the target state, $\hat{\rho}_{f}$, is a pure
state, the expression in Eq. \ref{eq:Fidelitet_mixade tillst=0000E5nd}
is reduced to

\begin{equation}
F(\hat{\rho})=\sqrt{\bra{N}\hat{\rho}\ket{N}}.
\end{equation}

\subsection{\label{sub:Non-Markovianity}Quantifying non-Markovianity}

A commonly used quantifier of non-Markovianity, based on distinguishability,
was introduced by Breuer, Laine and Piilo \citep{Breuer2009} and
is consequently referred to as the BLP-measure in literature. It is
derived from the observation that for all completely positive and
trace-preserving maps, $\Phi$, the following holds \citep{Ruskai1994};

\begin{equation}
D(\Phi(\hat{\rho}_{1}),\Phi(\hat{\rho}_{2}))\leq D(\hat{\rho}_{1},\hat{\rho}_{2}),
\end{equation}
where $D$ is the trace distance, defined as \citep{Nielsen}

\begin{equation}
D(\hat{\rho}_{1},\hat{\rho}_{2})=\frac{1}{2}{\rm {Tr}}\left|\hat{\rho}_{1}-\hat{\rho}_{2}\right|.
\end{equation}
Hence, the trace distance - which monitors the distinguishability
of two states - will be a monotonically decreasing function of time.
Any deviation from this behavior might be interpreted as a backflow
of information and hence, a non-Markovian time evolution. In \citep{Breuer2009},
the measure of non-Markovianity of a dynamical map, $\Phi$, is defined
as

\begin{equation}
\mathit{\mathrm{\mathcal{N}}}(\Phi)=\underset{\hat{\rho}_{1}(0),\hat{\rho}_{2}(0)}{\max}\underset{\sigma>0}{\int}\sigma[t,\hat{\rho}_{1}(0),\hat{\rho}_{2}(0)]\,dt,\label{eq:NM measure}
\end{equation}
where $\hat{\rho}_{1}(0)$ and $\hat{\rho}_{2}(0)$ are two initial
states and

\begin{equation}
\sigma[t,\hat{\rho}_{1}(0),\hat{\rho}_{2}(0)]=\frac{d}{dt}D(\hat{\rho}_{1},\hat{\rho}_{2}).
\end{equation}
The integral in Eq. \ref{eq:NM measure} is optimized over all possible
pairs of initial states to form a measure that only captures the features
of the dynamics. The physical and mathematical structure of optimal
state pairs is studied in Ref. \citep{Wissmann2012}. It is shown
that an optimal pair of initial states will be orthogonal (i.e., the
states can be described by density operators with orthogonal support),
which implies that such a state pair must belong to the boundary,
$\partial\mathcal{M}$, of the state space $\mathcal{M}$ of the density
operators describing states of the system. Optimal state-pairs will
fulfill $D(\hat{\rho}_{1}(0),\hat{\rho}_{2}(0))=1$, i.e., they are
initially completely distinguishable. In this study, we have used
the results in \citep{Wissmann2012} in the optimization procedure. 

In practice, we compute $\mathcal{N}$ by maximizing the integral
of $\sigma[t,\hat{\rho}_{1}(0),\hat{\rho}_{2}(0)]$ over a random
sample of initial state pairs. The orthogonality condition mentioned
above is taken into account by using that any pair of density matrices
$(\rho_{a},\rho_{b})$ with orthogonal support can be written as 
\begin{equation}
\left\{ \begin{array}{l}
\rho_{a}=U\tilde{\rho}_{a}U^{\dagger}\\
\rho_{b}=U\tilde{\rho}_{b}U^{\dagger}
\end{array}\right.,
\end{equation}
where $\tilde{\rho}_{a}$ and $\tilde{\rho}_{b}$ are density matrices
of the forms 
\begin{equation}
\tilde{\rho}_{a}=\mathrm{diag}(\lambda_{1},\ldots,\lambda_{m},0,\ldots,0)
\end{equation}
and 
\begin{equation}
\tilde{\rho}_{b}=\mathrm{diag}(0,\ldots,0,\lambda_{m+1},\ldots,\lambda_{N}),
\end{equation}
and $U$ is an element of the special unitary group of degree $N$,
SU($N$). Conversely, it also holds that any pair of matrices generated
according to the above prescription are density matrices with orthogonal
support. The set of all orthogonal state pairs of size $N$ can therefore
be parameterized by parametrizing SU($N$). In the present work, for
the $N=2$ case, we choose the parametrization 
\begin{equation}
U=\left[\begin{array}{cc}
\cos\theta e^{i\phi_{1}} & -\sin\theta e^{-i\phi_{2}}\\
\sin\theta e^{i\phi_{2}} & \cos\theta e^{-i\phi_{1}}
\end{array}\right],
\end{equation}

\begin{equation}
0\leq\theta\leq\pi/2,~0\leq\phi_{1},\phi_{2}\leq2\pi,
\end{equation}
whereas for the $N=3$ case we choose the analogous parametrization
(also in terms of angles $\theta_{i}$ and phases $\phi_{i}$) given
in Ref. \citep{Bronzan1988}. The sample of initial state pairs used
for evaluating $\mathcal{N}$ is then constructed by drawing $\lambda_{i}$,
$\theta_{i}$ and $\phi_{i}$ from a uniform distribution. Due to
the simplicity of the present approach (for example, we do not assure
that the distribution of initial state pairs is uniform) combined
with the rather small sample size ($10^{5}$ and $10^{4}$ pairs for
$N=2$ and $N=3$, respectively), we shall regard the values we obtain
for $\mathcal{N}$ only as estimates of the corresponding true values.

\subsection{\label{sub:Coherence}Quantifying coherence}

Following the work of Ref. \citep{Baumgratz2014}, the $l_{1}$-norm
of coherence is used to quantify the amount of coherence in the system.
For a system described by a density operator, $\hat{\rho}$, the $l_{1}$-norm
of coherence is given by

\begin{equation}
C_{l_{1}}^{(\psi)}(\hat{\rho})=\underset{\underset{i\neq j}{i,j}}{\sum}\left|\bra{\psi_{i}}\hat{\rho}\ket{\psi_{j}}\right|=2\underset{i<j}{\sum}\left|\rho_{ij}\right|,
\end{equation}

\noindent where $\ket{\psi_{i}}$ and $\ket{\psi_{j}}$ are two quantum
states in the set of the chosen basis states $\left\{ \ket{\psi_{i}}\right\} $. 

Since EET occurs in the site basis, coherence in the site basis, $C_{l_{1}}^{(s)}$,
is of interest. Especially, we study \textit{\textcolor{black}{local
coherence}}, 

\begin{equation}
C_{l_{1},ij}^{(s)}(\hat{\rho})=2\left|\rho_{ij}\right|,
\end{equation}
between sites $i$ and $j$, where $\ket{\psi_{i}}=\ket{i}$ and $\ket{\psi_{j}}=\ket{j}$.
Mathematically, the local coherence between two sites equals entanglement
as quantified by concurrence \citep{Wootters1998}, i.e., the amount
of entanglement between sites can be detected by this measure.

Furthermore, coherence in the exciton basis (the eigenbasis of the
system Hamiltonian), $C_{l_{1}}^{(e)}$, is investigated. In a closed
system, coherence in the exciton basis will be constant in time since
the excitons are stationary solution of the Schr\"odinger equation.
Hence, any time dependence of coherence in the exciton basis will
be due to environmental effects \textcolor{black}{and may be seen
as a in-and-out flow of information to the system.}

\section{Results and discussion}

In presenting our results, we use the following notations to describe
our findings:
\begin{itemize}
\item $F_{{\rm {OS,max}}}$$\left(F_{{\rm {OS,min}}}\right)$ denotes the
\textit{\textcolor{black}{maximal}} (\textit{\textcolor{black}{minimal}})
efficiency obtained for each dimer and trimer system when all environmental
parameters ($\lambda$, $\tau$ and $T$) as well as the time ($t$)
are optimized in the open system (OS). The corresponding optimal parameter
values are reported in the same manner, where for example $T_{{\rm {OS,max}}}$
is the temperature for which $F_{{\rm {OS,max}}}$ is achieved. When
the efficiency is optimized only over certain parameters, while the
dependencies on the others are \textcolor{black}{retained}, we show
the free parameters as a superscript in the already introduced notations.
For example, the efficiency optimized over $T$ and $t$ is denoted
$F_{{\rm {OS,max}}}^{(\lambda,\tau)}$while the efficiency optimized
over $\lambda$, $\tau$ and $T$ is denoted $F_{{\rm {OS,max}}}^{(t)}$. 
\item For the maximal (minimal) efficiency in the closed system (CS) (only
optimized over $t$) we use the notation $F_{{\rm {CS,max}}}$$\left(F_{{\rm {CS,min}}}\right)$.
The efficiency reached in the long-time limit (at equilibrium population),
as calculated by Eq. \ref{eq:Equilibrium}, is denoted $F_{{\rm {eq}}}$.
\item Coherence, $C_{l_{1}}^{(\psi)}$, is always shown for the optimal
values (with respect to efficiency) of $\lambda$, $\tau$ and $T$
and its time dependence will be shown explicitly. 
\item Non-Markovianity is reported for environmental parameter values corresponding
to maximal and minimal efficiency, and is denoted $\mathcal{N}_{{\rm {OS,max}}}$
and $\mathcal{N}_{{\rm {OS,min}}},$ respectively. For parameter values
corresponding to maximal efficiency, we report trace distance as $D_{{\rm {OS,max}}}$.
\end{itemize}
\noindent 
\begin{figure*}
\includegraphics[scale=0.2]{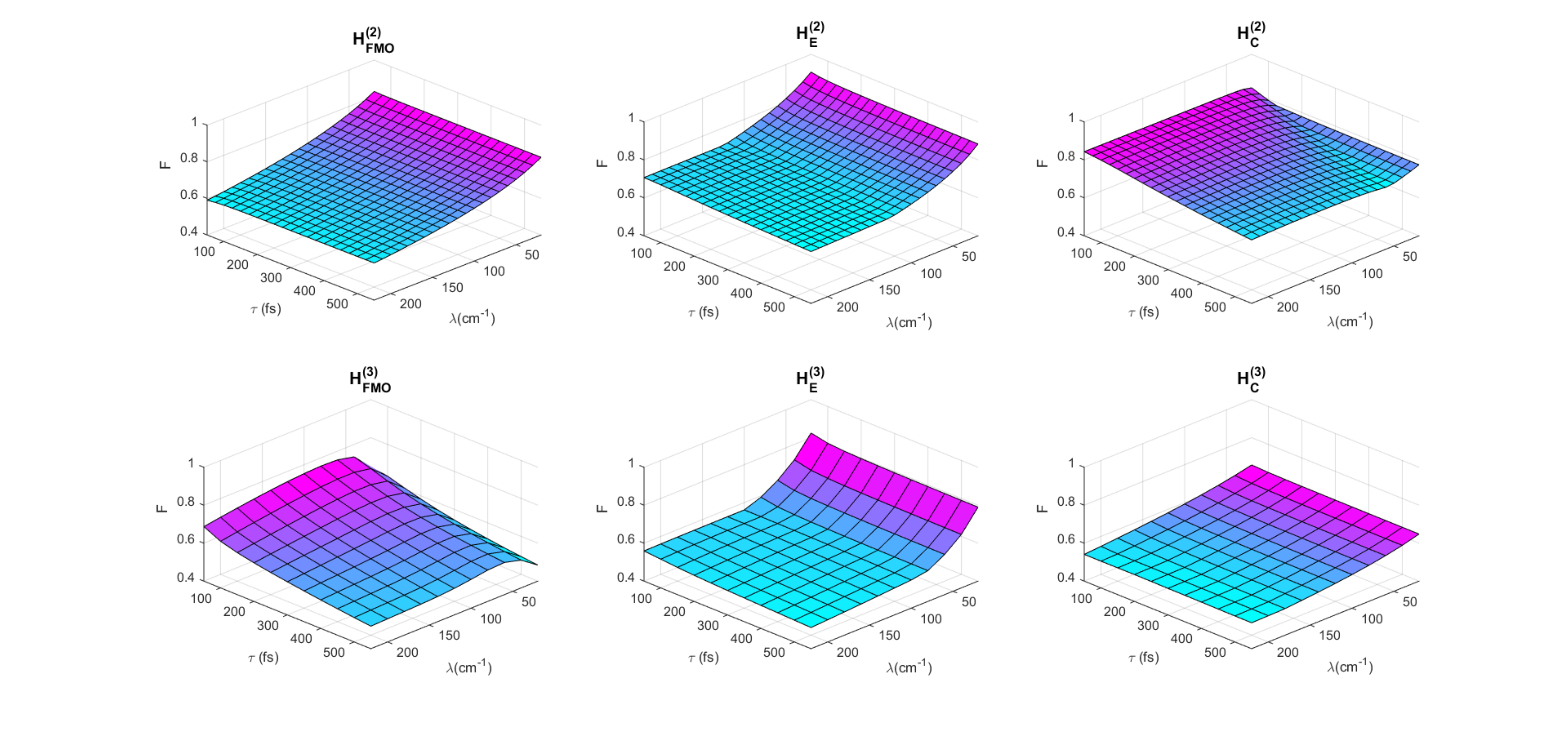}

\protect\caption{\label{fig:3D-plott}$F_{{\rm {OS,max}}}^{(\lambda,\tau)}$ as a function
of $\lambda$ and $\tau$ for $\hat{H}_{{\rm {FMO}}}^{(N)}$, $\hat{H}_{{\rm {\rm {E}}}}^{(N)}$
and $\hat{H}_{{\rm {C}}}^{(N)}$. The upper panel shows results for
the dimer systems and the lower panel shows results for the trimer
systems. }

\end{figure*}

\noindent 
\begin{figure*}
\includegraphics[scale=0.25]{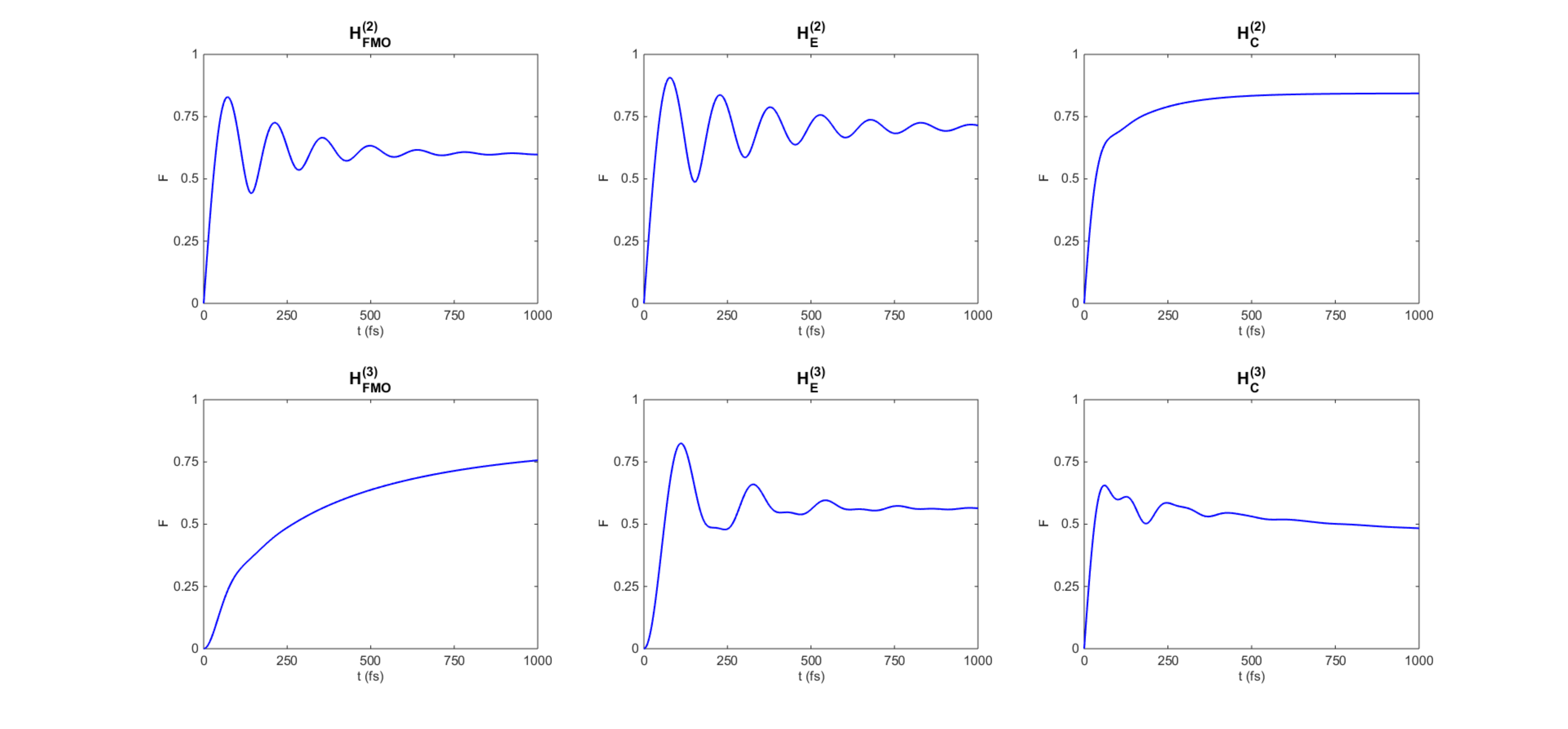}\protect\caption{\label{fig:F_OS}$F_{{\rm {OS,max}}}^{(t)}$ as a function of $t$
for $\hat{H}_{{\rm {FMO}}}^{(N)}$, $\hat{H}_{{\rm {E}}}^{(N)}$ and
$\hat{H}_{{\rm {C}}}^{(N)}$. The upper panel shows results for the
dimer systems and the lower panel shows results for the trimer systems. }
\end{figure*}

\noindent 
\begin{figure*}
\includegraphics[scale=0.25]{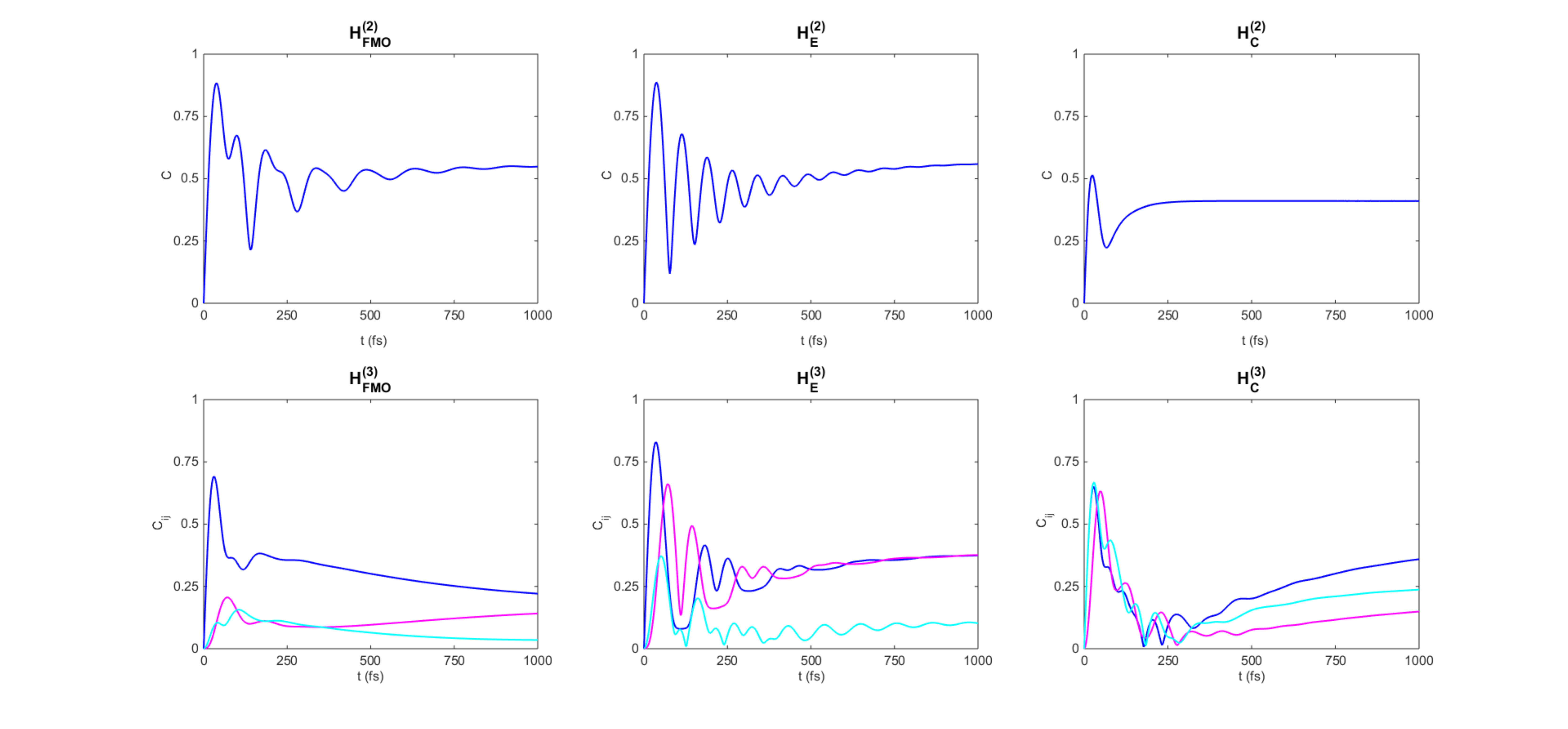}\protect\caption{\label{fig:C_site_OS}$C_{l_{1},ij}^{(s)}$ as a function of $t$
for $\hat{H}_{{\rm {FMO}}}^{(N)}$, $\hat{H}_{{\rm {\rm {E}}}}^{(N)}$
and $\hat{H}_{{\rm {C}}}^{(N)}$. The upper panel shows results for
the dimer systems and the lower panel shows results for the trimer
systems. Here, $C_{l_{1},12}^{(s)}$ (blue) $C_{l_{1},23}^{(s)}$
(magenta) and $C_{l_{1},13}^{(s)}$ (cyan) are shown.}

\end{figure*}

\noindent 
\begin{figure*}
\includegraphics[scale=0.25]{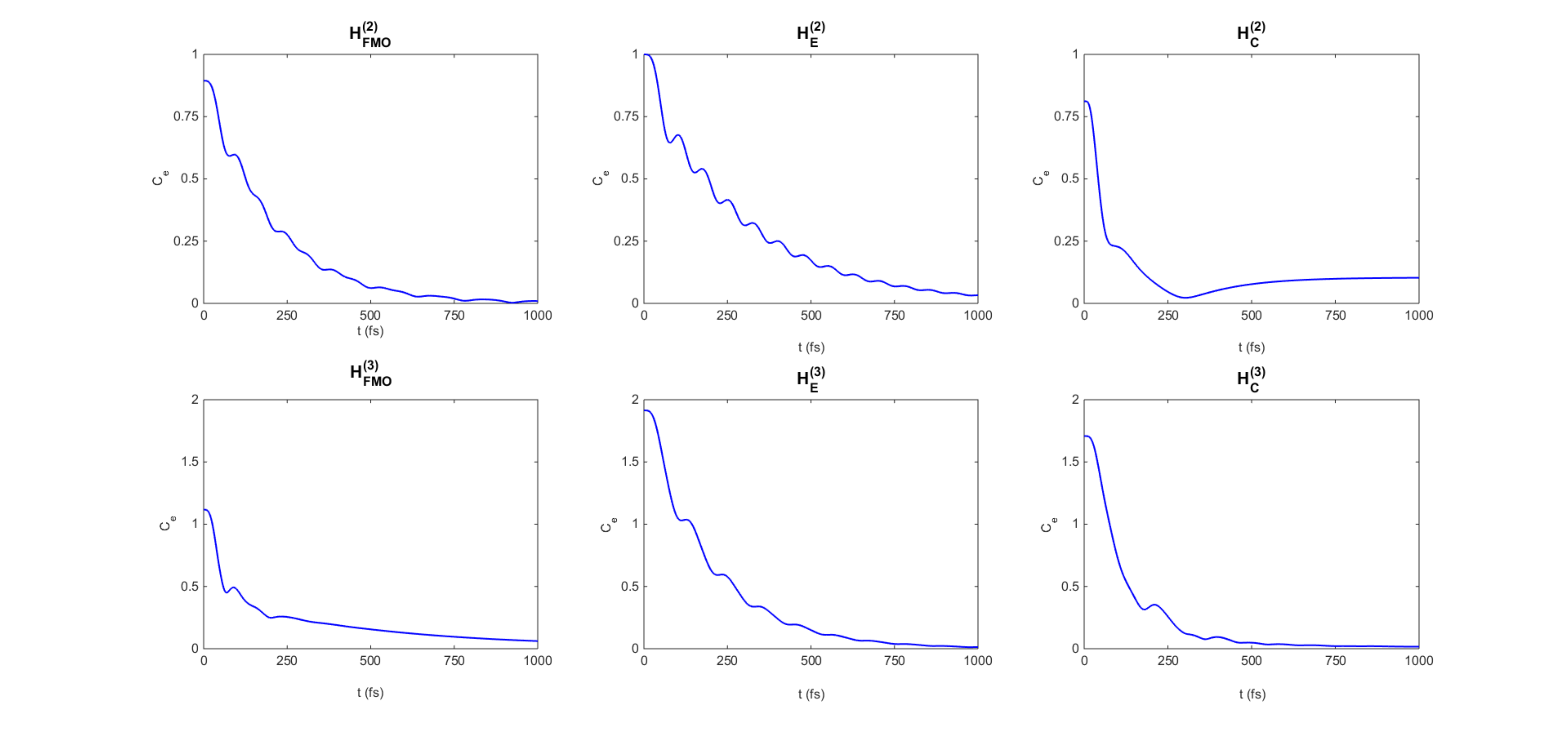}\protect\caption{\label{fig:C_exciton_OS}$C_{l_{1}}^{(e)}$ as a function of $t$
for $\hat{H}_{{\rm {FMO}}}^{(N)}$, $\hat{H}_{{\rm {\rm {E}}}}^{(N)}$
and $\hat{H}_{{\rm {C}}}^{(N)}$. The upper panel shows results for
the dimer systems and the lower panel shows results for the trimer
systems. }

\end{figure*}

\noindent The efficiency maximized over $T$ and $t$, i.e., $F_{{\rm {OS,max}}}^{(\lambda,\tau)}$,
is shown in Fig. \ref{fig:3D-plott} while the time dependence of
the efficiency maximized over all environmental parameters, $F_{{\rm {OS,max}}}^{(t)}$,
is shown in Fig. \ref{fig:F_OS}. \textcolor{black}{It should be noted
that in neither of the systems, the maximal efficiency shows a strong
temperature dependence, why we omit showing the temperature dependence
explicitly. }

The local coherences in the site basis, $C_{l_{1},ij}^{(s)}$, are
shown in Fig. \ref{fig:C_site_OS} while the coherence in the exciton
basis, $C_{l_{1}}^{(e)}$, is shown in Fig. \ref{fig:C_exciton_OS}.

\begin{figure*}
\includegraphics[scale=0.25]{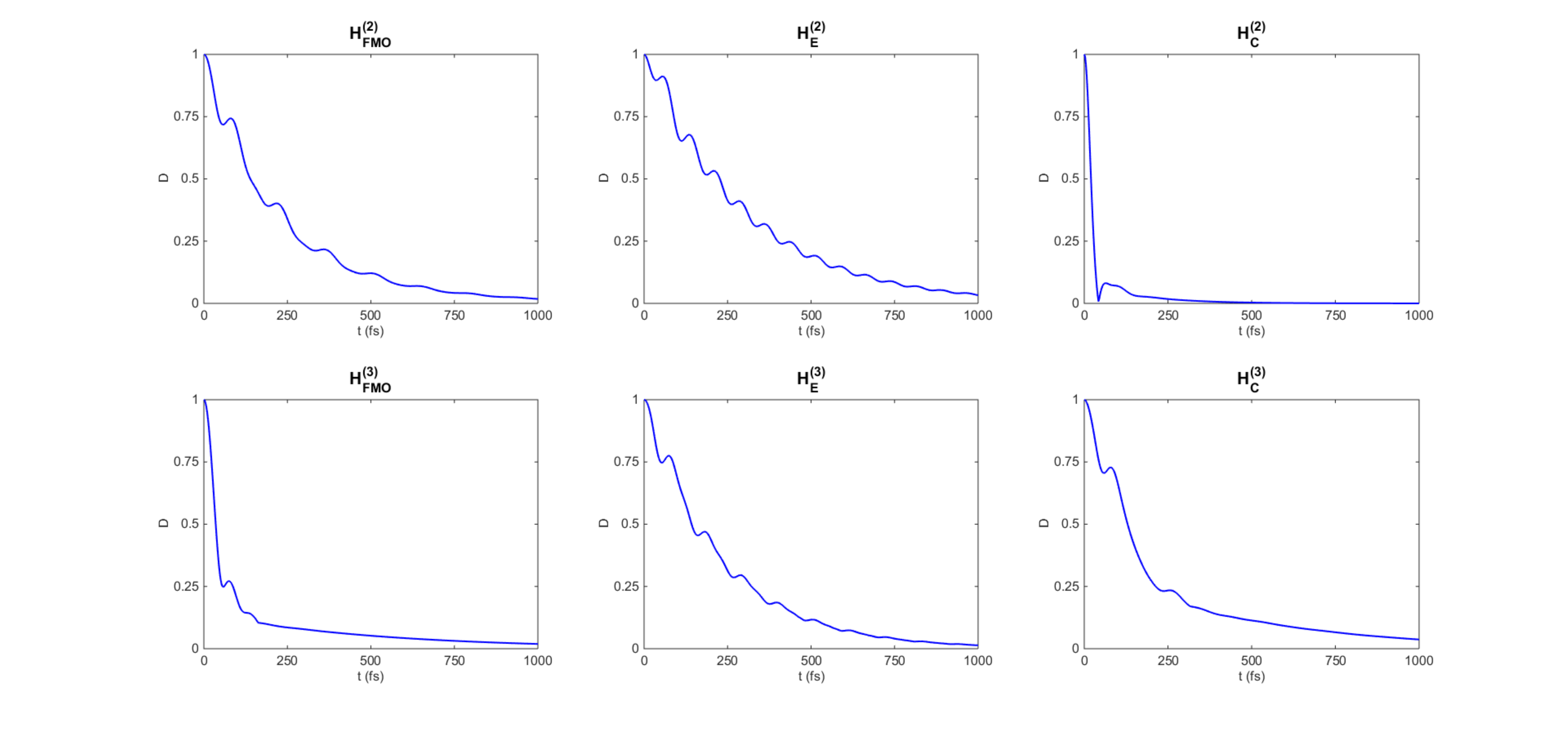}

\protect\caption{\label{fig:D_OS}$D_{{\rm {OS,max}}}$ as a function of $t$ for $\hat{H}_{{\rm {FMO}}}^{(N)}$,
$\hat{H}_{{\rm {E}}}^{(N)}$ and $\hat{H}_{{\rm {C}}}^{(N)}$. The
upper panel shows results for the dimer systems and the lower panel
shows results for the trimer systems. }
\end{figure*}

The value of the maximal efficiency, $F_{{\rm {OS,max}}}$, the values
of the environmental parameters and \textcolor{black}{the time} for
which $F_{{\rm {OS,max}}}$ is achieved, and the corresponding value
of non-Markovianity are given in Tab. \ref{tab:Dimer_F_max} for the
dimer systems and in Tab. \ref{tab:Trimer_F_max} for the trimer systems.
The time evolution of the trace distance of optimal state pairs, $D_{{\rm {OS,max}}}$,
is shown in Fig. \ref{fig:D_OS}. The values of the environmental
parameters and the corresponding value of non-Markovianity, but now
for minimal efficiency, $F_{{\rm {OS,min}}}$, are shown in Tab. \ref{tab:Dimer_F_min}
(dimers) and Tab. \ref{tab:Trimer_F_min} (trimers) for comparison.

\noindent 
\begin{table*}
\protect\caption{\label{tab:Dimer_F_max}Maximal efficiency in the open dimer systems,
$F_{{\rm {OS,max}}}$, environmental parameters and time for $F_{{\rm {OS,max}}}$,
and corresponding values of non-Markovianity. For comparison, maximal
efficiency in the closed system, $F_{{\rm {CS,max}}},$ and the efficiency
in the long-time limit, $F_{{\rm {eq}}},$ are also shown.}
\begin{tabular}{|c|c|c|c|c|c|c|c|c|}
\hline 
 & $F_{{\rm {OS,max}}}$ & $\lambda_{{\rm {OS,max}}}\:({\rm {cm}^{-1})}$ & $\tau_{{\rm {OS,max}}}\:({\rm {fs})}$ & $T_{{\rm {OS,max}}}\:({\rm {K})}$ & $t_{{\rm {OS,max}}}\:({\rm {fs})}$ & $\mathcal{N}_{{\rm {OS,max}}}$ & $F_{{\rm {CS,max}}}$ & $F_{{\rm {eq}}}$\tabularnewline
\hline 
\hline 
$\hat{H}_{{\rm {FMO}}}^{(2)}$ & $0.83$ & $20$ & $50$ & $250$ & $72.5$ & $0.045$ & $0.89$ & $0.61$\tabularnewline
\hline 
$\hat{H}_{{\rm {E}}}^{(2)}$ & $0.91$ & $20$ & $50$ & $250$ & $77.5$ & $0.11$ & $1$ & $0.71$\tabularnewline
\hline 
$\hat{H}_{{\rm {C}}}^{(2)}$ & $0.84$ & $220$ & $50$ & $250$ & $1000$ & $0.074$ & $0.81$ & $0.82$\tabularnewline
\hline 
\end{tabular}
\end{table*}

\noindent 
\begin{table*}
\protect\caption{\label{tab:Trimer_F_max}Maximal efficiency in the open trimer systems,
$F_{{\rm {OS,max}}}$, environmental parameters and time for $F_{{\rm {OS,max}}}$,
and corresponding values of non-Markovianity. For comparison, maximal
efficiency in the closed system, $F_{{\rm {CS,max}}},$ and the efficiency
in the long-time limit, $F_{{\rm {eq}}},$ are also shown.}
\begin{tabular}{|c|c|c|c|c|c|c|c|c|}
\hline 
 & $F_{{\rm {OS,max}}}$ & $\lambda_{{\rm {OS,max}}}\:({\rm {cm}^{-1})}$ & $\tau_{{\rm {OS,max}}}\:({\rm {fs})}$ & $T_{{\rm {OS,max}}}\:({\rm {K})}$ & $t_{{\rm {OS,max}}}\:({\rm {fs})}$ & $\mathcal{N}_{{\rm {OS,max}}}$ & $F_{{\rm {CS,max}}}$ & $F_{{\rm {eq}}}$\tabularnewline
\hline 
\hline 
$\hat{H}_{{\rm {FMO}}}^{(3)}$ & $0.76$ & $80$ & $50$ & $250$ & $1000$ & $0.024$ & $0.29$ & $0.80$\tabularnewline
\hline 
$\hat{H}_{{\rm {E}}}^{(3)}$ & $0.82$ & $20$ & $50$ & $250$ & $110$ & $0.069$ & $1$ & $0.56$\tabularnewline
\hline 
$\hat{H}_{{\rm {C}}}^{(3)}$ & $0.65$ & $20$ & $50$ & $250$ & $60$ & $0.026$ & $0.71$ & $0.48$\tabularnewline
\hline 
\end{tabular}
\end{table*}

\noindent 

\noindent 
\begin{table*}
\protect\caption{\label{tab:Dimer_F_min}Minimal efficiency in the open dimer systems,
$F_{{\rm {OS,min}}}$, environmental parameters for $F_{{\rm {OS,min}}}$
and corresponding values of non-Markovianity.}
\begin{tabular}{|c|c|c|c|c|c|}
\hline 
 & $F_{{\rm {OS,min}}}$ & $\lambda_{{\rm {OS,min}}}\:({\rm {cm}^{-1})}$ & $\tau_{{\rm {OS,min}}}\:({\rm {fs})}$ & $T_{{\rm {OS,min}}}\:({\rm {K}})$ & $\mathcal{N_{{\rm {OS,min}}}}$\tabularnewline
\hline 
\hline 
$\hat{H}_{{\rm {FMO}}}^{(2)}$ & $0.59$ & $220$ & $50$ & $250$ & $0.079$\tabularnewline
\hline 
$\hat{H}_{{\rm {E}}}^{(2)}$ & $0.67$ & $220$ & $550$ & $250$ & $0.64$\tabularnewline
\hline 
$\hat{H}_{{\rm {C}}}^{(2)}$ & $0.72$ & $50$ & $550$ & $300$ & $0.76$\tabularnewline
\hline 
\end{tabular}

\end{table*}

\noindent 
\begin{table*}
\protect\caption{\label{tab:Trimer_F_min}Minimal efficiency in the open trimer systems,
$F_{{\rm {OS,min}}}$, environmental parameters for $F_{{\rm {OS,min}}}$
and corresponding values of non-Markovianity.}
\begin{tabular}{|c|c|c|c|c|c|}
\hline 
 & $F_{{\rm {OS,min}}}$ & $\lambda_{{\rm {OS,min}}}\:({\rm {cm}^{-1})}$ & $\tau_{{\rm {OS,min}}}\:({\rm {fs})}$ & $T_{{\rm {OS,min}}}\:({\rm {K}})$ & $\mathcal{N_{{\rm {OS,min}}}}$\tabularnewline
\hline 
\hline 
$\hat{H}_{{\rm {FMO}}}^{(3)}$ & $0.48$ & $20$ & $550$ & $250$ & $0.50$\tabularnewline
\hline 
$\hat{H}_{{\rm {E}}}^{(3)}$ & $0.51$ & $220$ & $550$ & $300$ & $0.37$\tabularnewline
\hline 
$\hat{H}_{{\rm {C}}}^{(3)}$ & $0.54$ & $220$ & $550$ & $250$ & $0.30$\tabularnewline
\hline 
\end{tabular}
\end{table*}

\subsection{\label{sub:Dimer}Dimer}

As can be seen in Tab. \ref{tab:Dimer_F_max}, EET from site $1$
to $2$ in the systems with Hamiltonians $\hat{H}_{{\rm {FMO}}}^{(2)}$
and $\hat{H}_{{\rm {E}}}^{(2)}$ is optimal for the same values of
the environmental parameters, which are the lowest possible in the
defined regimes and hence yields EET close to a fully coherent evolution
(due to the low value of $\lambda_{{\rm {OS,max}}}$). \textcolor{black}{For
the system $\hat{H}_{{\rm {C}}}^{(2)}$, the scenario is different.
Here, $\lambda_{{\rm {OS,max}}}$ is as large as possible in the defined
regime (the optimal environmental values for $T$ and $\tau$ coincide
with $\hat{H}_{{\rm {FMO}}}^{(2)}$ and $\hat{H}_{{\rm {E}}}^{(2)}$)
and should be associated with an incoherent evolution. We discuss
the optimal environmental regimes for the different systems in more
detail below. }
\begin{itemize}
\item \textcolor{black}{For the system with Hamiltonian $\hat{H}_{{\rm {FMO}}}^{(2)},$
$E_{2}>E_{1}$ and at equilibrium site $1$ will be the most populated.
Hence, EET from site $1$ to site $2$ may benefit from an EET within
the coherent regime since it could allow for quantum tunneling over
the energy barrier between the sites. Indeed, from the numerical calculations
it is found that $F_{{\rm {CS,max}}}\approx\frac{3}{2}F_{{\rm {eq}}}$,
which confirms that the most important feature of the environment
is to preserve coherence. }In Fig. \ref{fig:C_site_OS} it can be
seen that a large amount of coherence between site $1$ and $2$ is
built up on a short time scale. Fig. \ref{fig:C_exciton_OS} reveals
a monotonically decreasing, oscillating coherence in the exciton basis.
A similar behavior can be seen for $D_{{\rm {OS,max}}}$ in Fig. \ref{fig:D_OS}.
\item For the system with Hamiltonian $\hat{H}_{{\rm {E}}}^{(2)}$, $F_{{\rm {CS,max}}}=1$
(see Ref. \citep{Bengtson2017}) and $E_{1}=E_{2}.$ In this case,
$F_{{\rm {eq}}}=0.71$ and $F_{{\rm {OS,min}}}=0.67$ for $\lambda_{{\rm {OS,min}}}=220\:{\rm {cm}^{-1}}.$
Again, as in the case of $\hat{H}_{{\rm {FMO}}}^{(2)},$ the EET is
favored by environmental interactions that preserve a coherent EET,
which implies preservation of coherence \textcolor{black}{in the site
basis (see Fig. \ref{fig:C_site_OS}). Here, $C_{l_{1}}^{(e)}$ and
$D_{{\rm {OS,max}}}$ follows the same pattern as for $\hat{H}_{{\rm {FMO}}}^{(2)}$.}
\item In the last dimer system, with system Hamiltonian $\hat{H}_{{\rm {C}}}^{(2)},$
the situation is different - as already noted - from $\hat{H}_{{\rm {FMO}}}^{(2)}$
and $\hat{H}_{{\rm {E}}}^{(2)}$, which can be seen in Tab. \ref{tab:Dimer_F_max}
and Fig. \ref{fig:F_OS}. Still, even though $F_{{\rm {OS,max}}}$
in this case (unlike the others) is achieved for a large value of
$\lambda$ ($\lambda_{{\rm {OS,max}}}=220\:{\rm {cm}^{-1}}$) , the
dependence on $\lambda$ is not very strong (see Fig. \ref{fig:3D-plott}).
From Tab. \ref{tab:Dimer_F_max} it can further be seen that the EET can occur with approximately equal efficiency in the very
coherent regime ($F_{{\rm {CS,max}}}=0.81$ ) as in the long time
limit ($F_{{\rm {eq}}}=0.82$). Hence, efficient EET could be expected
for either a small value of $\lambda$ or a small value of $\tau$.
\textcolor{black}{It should also be noted that $F_{{\rm {OS,max}}}$
($0.84$) does not seem to approach $F_{{\rm {eq}}}$ in the long-time
limit. In Ref. \citep{Subasi2012} it is shown that the Boltzmann
population does not describe the system equilibrium when the system
and the environment is strongly coupled, which might be the reason
why $F_{{\rm {OS,max}}}$ is even higher than $F_{{\rm {eq}}}$ according
to Eq. \ref{eq:Equilibrium} in this case. The coherence in the exciton
basis, as shown in Fig. \ref{fig:C_exciton_OS}, reveals that indeed
does the system approach a steady state that differs from the Boltzmann
state since the coherence approaches a stationary non-zero value.
Another interesting feature of the system $\hat{H}_{{\rm {C}}}^{(2)}$
is that even though oscillations in the coherence in the exciton basis
(which might indicate information back flow from the environment)
are quenched at a time scale of less than $100$ ${\rm {fs}}$, the
value of $\mathcal{N}_{{\rm {OS,max}}}$ is higher than for $\hat{H}_{{\rm {FMO}}}^{(2)}$.
This despite that the coherence in the exciton basis is maintained
for much longer in the latter case.} The information backflow in \textcolor{black}{$\hat{H}_{{\rm {C}}}^{(2)}$}
consists of one single peak, arising from zero distinguishability,
which can be seen in Fig. \ref{fig:D_OS}. Since non-Markovianity
is zero when the initial states for example are chosen as $\hat{\rho}_{1}(0)=\ket{1}\bra{1}$
and $\hat{\rho}_{2}(0)=\ket{2}\bra{2}$, it also shows the necessity
of using the non-Markovian measure in the way it is intended to and
not only select one or a few pairs of initial states. 
\end{itemize}
\textcolor{black}{By comparing the values of $F_{{\rm {OS,max}}}$
for each system in Tab. \ref{tab:Dimer_F_max} to the values of $F_{{\rm {OS,min}}}$
in Tab. \ref{tab:Dimer_F_min} together with the corresponding values
of non-Markovianity in each case, it can be seen that non-Markovianty
is higher for environments corresponding to minimal efficiency in
all three dimer systems. Hence it can be concluded that non-Markovianity
does not seem to be a resource - in the sense that a higher value
of non-Markovianity correlates with a higher EET efficiency - for
efficient EET in these dimer systems. The results also indicate that,
as could have been expected, $\tau$ is the parameter that mainly
governs the amount of non-Markovianity.}

\subsection{\label{sub:Trimer}Trimer}

\noindent In the trimer case, EET from site $1$ to $3$ is optimal
for the same values of the environmental parameters for both the system
with Hamiltonian $\hat{H}_{{\rm {E}}}^{(3)}$ and the system with
Hamiltonian $\hat{H}_{{\rm {C}}}^{(3)}$, which are the lowest possible
in the defined regimes as can be seen in Tab. \ref{tab:Trimer_F_max}.
Here, it is the system with Hamiltonian $\hat{H}_{{\rm {FMO}}}^{(3)}$
that is optimized in a less coherent regime than the other two systems.
\textcolor{black}{We discuss these results in more detail below. }
\begin{itemize}
\item \textcolor{black}{The values of the optimal environmental parameters
for the system with Hamiltonian $\hat{H}_{{\rm {FMO}}}^{(3)}$ are
such that $T_{{\rm {OS,max}}}$ and $\tau_{{\rm {OS,max}}}$ are the
lowest possible, but $\lambda_{{\rm {OS,max}}}$ is in an intermediate
regime ($80\:{\rm {cm^{-1}}})$. For this system, $F_{{\rm {CS,max}}}$
is very small ($0.29$) and consequently, $F_{{\rm {OS,min}}}$ is
obtained at $\lambda_{{\rm {OS,min}}}=20\:{\rm {cm}^{-1}}$. In Fig.
\ref{fig:3D-plott} it can be seen that for $\lambda>80\:{\rm {cm}^{-1}}$
the efficiency is also considerably reduced and in Tab. \ref{tab:Trimer_F_max}
it can be noted that $F_{{\rm {eq}}}>F_{{\rm {OS,max}}}$, i.e., the
efficiency in the long-time limit is higher than the value achieved
at optimized parameter values. Here seems to be a case where the time
scale of EET comes into play. Reaching equilibrium population for
an initial population of site $2$ could be expected to be a faster
process than from site $1$ since $E_{2}>E_{1}>E_{3}$. Hence, EET
may benefit from - at least partly - taking the route over site $2$.
A mechanism where the excitation can be transferred from site $1$
to $2$, despite the energy barrier, is quantum tunneling, i.e., a
coherent dynamics initially can transfer the excitation to site $2$
and then, from there, an energetic downhill funneling is most efficient
to transfer the excitation to site $3$. In Fig. \ref{fig:C_site_OS}
it can be seen that a large amount of coherence is built up between
site $1$ and $2$ on a short time scale. The maximum of EET efficiency
(with respect to $\lambda$ and $\tau$), observed in Fig. \ref{fig:3D-plott}
can be understood as an interplay between coherent dynamics to transfer
the excitation from site $1$ to $2$ and quenching of the coherent
dynamics when the population is located at site $2$ to break reversibility.
In the given time frame, this seems to be the most efficient EET process
for this system. }
\item \textcolor{black}{The system with Hamiltonian $\hat{H}_{{\rm {E}}}^{(3)}$,
with $F_{{\rm {CS,max}}}=1$, follows the same pattern as the corresponding
system in the dimer case. Indeed are the values of the optimal environmental
parameters as low as possible in the defined regime and provide a
close-to-coherent EET dynamics.}
\item \textcolor{black}{The last trimer system, with system Hamiltonian
$\hat{H}_{{\rm {C}}}^{(3)}$, has a structure where $E_{1}>E_{3}>E_{2}$
and the energy barrier between site $2$ and $3$ is large. At the
same time, the coupling between these two sites is very weak in comparison
to the energy difference between them as well as to the coupling to
site $1$. Equilibrium population naturally favors site $2$ over
site $3$ and therefore $F_{{\rm {eq}}}$} is rather small ($0.48$)\textcolor{black}{.
In the closed system however, $F_{{\rm {CS,max}}}=0.71$ and hence,
this system benefits from a coherent EET dynamics. }
\end{itemize}
\textcolor{black}{The coherence in the exciton basis is monotonically
decaying, more or less oscillating, for all trimer systems. For the
system $\hat{H}_{{\rm {FMO}}}^{(3)}$, the pattern is initially similar
to $\hat{H}_{{\rm {C}}}^{(2)}$and the other two trimer systems resembles
$\hat{H}_{{\rm {\rm {FMO}}}}^{(2)}$ and $\hat{H}_{{\rm {E}}}^{(2)}$.
Again, there is a similarity between the appearance of $C_{l_{1}}^{(e)}$
and $D_{{\rm {OS,max}}}$. }

\textcolor{black}{As in the dimer systems, there does not seem to
be any indications that non-Markovianity would be a resource for efficient
EET. By comparing $\mathcal{N}_{{\rm {OS,max}}}$ to $\mathcal{N}_{{\rm {OS,min}}}$
and the corresponding parameter values in Tabs. \ref{tab:Trimer_F_max}
and \ref{tab:Trimer_F_min}, the observation that a larger value of
$\tau$ yields a larger value of $\mathcal{N}$ is confirmed.}

\section{Conclusions}

\textcolor{black}{Previous studies on environment-assisted quantum
transport (ENAQT) in pigment-protein molecular aggregates }\citep{Qin2014,Plenio2008,Caruso2009,Chin2010,Sinayskiy2012,Marais2013,Wu2010,Rebentrost2009_1,Rebentrost2009_2,Mohseni2008,Dijkstra2012,Shabani2012,Mohseni2014}\textcolor{black}{{}
have, among others, proposed the following mechanisms for an enhanced
EET efficiency:}
\begin{itemize}
\item \textcolor{black}{The environment can suppress pathways not leading
to an enhanced probability of finding the excitation on the end site.}
\item \textcolor{black}{Strongly coupled vibrational modes in the environment
can induce resonant energy transfer between two sites/excitons that
in the static picture are non-resonant.}
\end{itemize}
\textcolor{black}{In particular it has been noted that for EET that
is very inefficient in the coherent regime, as the EET from site $1$
to site $3$ in the FMO-complex monomer, the efficiency can be increased
considerably by adding environmental effects. In this study, we do
not aim to explain the exact mechanisms for ENAQT, but rather try
to pin down some simple design principles (in terms of environmental
properties) for how to achieve efficient EET in model pigment-protein
molecular aggregates. }

\textcolor{black}{Our study indicates that the information about optimal
environmental conditions for efficient EET is in the structure of
the closed system Hamiltonian and in particular, the relationship
between the site energies. In general, a strategy to predict the optimal
environment for a system is to compare $F_{{\rm {CS,max}}}$ to $F_{{\rm {eq}}}$
of site $N$. If $F_{{\rm {CS,max}}}>F_{{\rm {eq}}}$, the most beneficial
role of the environment to promote efficient EET is to preserve coherent
EET dynamics (i.e., the system-environment coupling, $\lambda,$ should
be small). If instead $F_{{\rm {eq}}}>F_{{\rm {CS,max}}}$, most efficient
EET is achieved when the the system is driven towards equilibrium
as fast as possible (i.e., the environmental time scale, $\tau$,
should be small). In this study, efficient EET hence occurs in either
of the two regimes ``coherent dynamics'' (as close to the closed
system dynamics as possible) or ``equilibrating'' (as close to the
equilibrium population as possible). }

\textcolor{black}{In terms of the system Hamiltonian parameter space,
our results indicate that for systems where site $N$ is higher in
energy than site $1$, the EET benefits from quantum tunneling between
the sites and hence requires a dynamics within the coherent regime.
The same is true for the system Hamiltonians where $E_{N}=E_{1}.$
In such a case, the most important design principle is to create an
environment that allows for coherent EET dynamics to persist, i.e.,
the system should be weakly coupled to the environment. The environmental
time scale is less important. If instead $E_{1}>E_{N}$, the system
is favoring a fast return to equilibrium where $E_{N}$ obviously
is more populated than $E_{1}$. Here, the environmental time scale
should be designed to be as small as possible and the coupling strength
between the system and the environment is of less importance. Still,
depending on the energy landscape of the other sites in the network
as well as their intersite couplings, the time scale of equilibration
might be beyond the given time frame. Hence, in reality the most efficient
EET process might be a combination of coherent tunneling between sites
and a downhill funneling from site to site. We believe that the EET
in the system with Hamiltonian $\hat{H}_{{\rm {FMO}}}^{(3)}$ is an
example of such a combined process. }

\textcolor{black}{Previous studies on non-Markovian effects in pigment-protein
molecular aggregates have mainly connected the occurrence of non-Markovianity
to a prolonged duration of coherence \citep{Caruso2010,Chen2014,Thorwart2009}.
In \citep{Rebentrost2011} it is noted though that in a FMO complex
monomer, maximal non-Markovianity is within the same environmental
regime as optimal EET efficiency according to Ref. \citep{Rebentrost2009_1}.
The authors suggest that the role of non-Markovianity could be to
preserve coherence, which could be of importance for the first EET
route (site $1$, $2$ and $3$) since there is an energy barrier
between site $1$ and $2$ (which might require quantum tunneling).
Even though the important features of the system Hamiltonian parameter
space for these three sites coincide with our trimer system with Hamiltonian
$\hat{H}_{{\rm {FMO}}}^{(3)}$ (with only the exact parameter values
differing), the dynamics will differ a bit due to the omission of
the other four pigments in the FMO-complex monomer in our calculations.
It should be noted that even though non-Markovianity is quantified
by the same measure as in the present study, the maximization over
initial states is ignored. This might affect the conclusion about
the connection between non-Markovianity and EET efficiency. Indeed
our study does not indicate a general connection between EET efficiency
and non-Markovianity. }

\textcolor{black}{The parameter that seemingly affects the amount
of non-Markovianity - and hence, the backflow of information from
the environment to the system - the most, is the environmental time
scale $(\tau$). The systems where $F_{{\rm {OS,max}}}$ is achieved
in the coherent regime are not particularly sensitive to a potential
backflow of information as accompanied with large values of $\tau$,
but they do not seem to benefit from it either.}

Another general feature worth noticing is that the pattern of the trace
distance is very similar to the pattern of coherence in the exciton
basis when $\lambda$ is small (see Figs. \ref{fig:C_exciton_OS}
and \ref{fig:D_OS}). This result suggests that a temporal increase
of coherence in the exciton basis indeed may be interpreted as a backflow
of information from the environment to the system and could possibly
detect non-Markovian dynamics.

Lastly, our study shows the necessity of using the BLP-measure the way
that it is intended to since a predefined choice of an initial state
pair may give an arbitrary result.

\section*{Acknowledgments}

The computations were performed on resources provided by the Swedish
National Infrastructure for Computing (SNIC) at Uppsala Multidisciplinary
Center for Advanced Computational Science (UPPMAX) under Project snic2017-7-140. 

\bibliographystyle{apsrev4-1}
\bibliography{Referensbibliotek}

\begin{thebibliography}{58}%
\makeatletter
\providecommand \@ifxundefined [1]{%
 \@ifx{#1\undefined}
}%
\providecommand \@ifnum [1]{%
 \ifnum #1\expandafter \@firstoftwo
 \else \expandafter \@secondoftwo
 \fi
}%
\providecommand \@ifx [1]{%
 \ifx #1\expandafter \@firstoftwo
 \else \expandafter \@secondoftwo
 \fi
}%
\providecommand \natexlab [1]{#1}%
\providecommand \enquote  [1]{``#1''}%
\providecommand \bibnamefont  [1]{#1}%
\providecommand \bibfnamefont [1]{#1}%
\providecommand \citenamefont [1]{#1}%
\providecommand \href@noop [0]{\@secondoftwo}%
\providecommand \href [0]{\begingroup \@sanitize@url \@href}%
\providecommand \@href[1]{\@@startlink{#1}\@@href}%
\providecommand \@@href[1]{\endgroup#1\@@endlink}%
\providecommand \@sanitize@url [0]{\catcode `\\12\catcode `\$12\catcode
  `\&12\catcode `\#12\catcode `\^12\catcode `\_12\catcode `\%12\relax}%
\providecommand \@@startlink[1]{}%
\providecommand \@@endlink[0]{}%
\providecommand \url  [0]{\begingroup\@sanitize@url \@url }%
\providecommand \@url [1]{\endgroup\@href {#1}{\urlprefix }}%
\providecommand \urlprefix  [0]{URL }%
\providecommand \Eprint [0]{\href }%
\providecommand \doibase [0]{http://dx.doi.org/}%
\providecommand \selectlanguage [0]{\@gobble}%
\providecommand \bibinfo  [0]{\@secondoftwo}%
\providecommand \bibfield  [0]{\@secondoftwo}%
\providecommand \translation [1]{[#1]}%
\providecommand \BibitemOpen [0]{}%
\providecommand \bibitemStop [0]{}%
\providecommand \bibitemNoStop [0]{.\EOS\space}%
\providecommand \EOS [0]{\spacefactor3000\relax}%
\providecommand \BibitemShut  [1]{\csname bibitem#1\endcsname}%
\let\auto@bib@innerbib\@empty
\bibitem [{\citenamefont {Breuer}\ and\ \citenamefont
  {Petruccione}(2002)}]{Breuer}%
  \BibitemOpen
  \bibfield  {author} {\bibinfo {author} {\bibfnamefont {H.-P.}\ \bibnamefont
  {Breuer}}\ and\ \bibinfo {author} {\bibfnamefont {F.}~\bibnamefont
  {Petruccione}},\ }\href@noop {} {\emph {\bibinfo {title} {The theory of open
  quantum systems}}}\ (\bibinfo  {publisher} {Oxford University Press},\
  \bibinfo {year} {2002})\BibitemShut {NoStop}%
\bibitem [{\citenamefont {Engel}\ \emph {et~al.}(2007)\citenamefont {Engel},
  \citenamefont {Calhoun}, \citenamefont {Read}, \citenamefont {Ahn},
  \citenamefont {Mancal}, \citenamefont {Cheng}, \citenamefont {Blankenship},\
  and\ \citenamefont {Fleming}}]{Engel2007}%
  \BibitemOpen
  \bibfield  {author} {\bibinfo {author} {\bibfnamefont {G.~S.}\ \bibnamefont
  {Engel}}, \bibinfo {author} {\bibfnamefont {T.~R.}\ \bibnamefont {Calhoun}},
  \bibinfo {author} {\bibfnamefont {E.~L.}\ \bibnamefont {Read}}, \bibinfo
  {author} {\bibfnamefont {T.-K.}\ \bibnamefont {Ahn}}, \bibinfo {author}
  {\bibfnamefont {T.}~\bibnamefont {Mancal}}, \bibinfo {author} {\bibfnamefont
  {Y.-C.}\ \bibnamefont {Cheng}}, \bibinfo {author} {\bibfnamefont {R.~E.}\
  \bibnamefont {Blankenship}}, \ and\ \bibinfo {author} {\bibfnamefont {G.~R.}\
  \bibnamefont {Fleming}},\ }\href@noop {} {\bibfield  {journal} {\bibinfo
  {journal} {Nature}\ }\textbf {\bibinfo {volume} {446}},\ \bibinfo {pages}
  {782} (\bibinfo {year} {2007})}\BibitemShut {NoStop}%
\bibitem [{\citenamefont {Hayes}\ \emph {et~al.}(2010)\citenamefont {Hayes},
  \citenamefont {Panitchayangkoon}, \citenamefont {Fransted}, \citenamefont
  {Caram}, \citenamefont {Wen}, \citenamefont {Freed},\ and\ \citenamefont
  {Engel}}]{Hayes2010}%
  \BibitemOpen
  \bibfield  {author} {\bibinfo {author} {\bibfnamefont {D.}~\bibnamefont
  {Hayes}}, \bibinfo {author} {\bibfnamefont {G.}~\bibnamefont
  {Panitchayangkoon}}, \bibinfo {author} {\bibfnamefont {K.~A.}\ \bibnamefont
  {Fransted}}, \bibinfo {author} {\bibfnamefont {J.~R.}\ \bibnamefont {Caram}},
  \bibinfo {author} {\bibfnamefont {J.}~\bibnamefont {Wen}}, \bibinfo {author}
  {\bibfnamefont {K.~F.}\ \bibnamefont {Freed}}, \ and\ \bibinfo {author}
  {\bibfnamefont {G.~S.}\ \bibnamefont {Engel}},\ }\href@noop {} {\bibfield
  {journal} {\bibinfo  {journal} {New J. Phys.}\ }\textbf {\bibinfo {volume}
  {12}},\ \bibinfo {pages} {065042} (\bibinfo {year} {2010})}\BibitemShut
  {NoStop}%
\bibitem [{\citenamefont {Panitchayangkoon}\ \emph {et~al.}(2010)\citenamefont
  {Panitchayangkoon}, \citenamefont {Hayes}, \citenamefont {Fransted},
  \citenamefont {Caram}, \citenamefont {Harel}, \citenamefont {Wen},
  \citenamefont {Blankenship},\ and\ \citenamefont
  {Engel}}]{Panitchayangkoon2010}%
  \BibitemOpen
  \bibfield  {author} {\bibinfo {author} {\bibfnamefont {G.}~\bibnamefont
  {Panitchayangkoon}}, \bibinfo {author} {\bibfnamefont {D.}~\bibnamefont
  {Hayes}}, \bibinfo {author} {\bibfnamefont {K.~A.}\ \bibnamefont {Fransted}},
  \bibinfo {author} {\bibfnamefont {J.~R.}\ \bibnamefont {Caram}}, \bibinfo
  {author} {\bibfnamefont {E.}~\bibnamefont {Harel}}, \bibinfo {author}
  {\bibfnamefont {J.}~\bibnamefont {Wen}}, \bibinfo {author} {\bibfnamefont
  {R.~E.}\ \bibnamefont {Blankenship}}, \ and\ \bibinfo {author} {\bibfnamefont
  {G.~S.}\ \bibnamefont {Engel}},\ }\href@noop {} {\bibfield  {journal}
  {\bibinfo  {journal} {Proc. Natl. Acad. Sci. U.S.A.}\ }\textbf {\bibinfo
  {volume} {107}},\ \bibinfo {pages} {12766} (\bibinfo {year}
  {2010})}\BibitemShut {NoStop}%
\bibitem [{\citenamefont {Panitchayangkoon}\ \emph {et~al.}(2011)\citenamefont
  {Panitchayangkoon}, \citenamefont {Voronine}, \citenamefont {Abramavicius},
  \citenamefont {Caram}, \citenamefont {Lewis}, \citenamefont {Mukamel},\ and\
  \citenamefont {Engel}}]{Panitchayangkoon2011}%
  \BibitemOpen
  \bibfield  {author} {\bibinfo {author} {\bibfnamefont {G.}~\bibnamefont
  {Panitchayangkoon}}, \bibinfo {author} {\bibfnamefont {D.~V.}\ \bibnamefont
  {Voronine}}, \bibinfo {author} {\bibfnamefont {D.}~\bibnamefont
  {Abramavicius}}, \bibinfo {author} {\bibfnamefont {J.~R.}\ \bibnamefont
  {Caram}}, \bibinfo {author} {\bibfnamefont {N.~H.}\ \bibnamefont {Lewis}},
  \bibinfo {author} {\bibfnamefont {S.}~\bibnamefont {Mukamel}}, \ and\
  \bibinfo {author} {\bibfnamefont {G.~S.}\ \bibnamefont {Engel}},\ }\href@noop
  {} {\bibfield  {journal} {\bibinfo  {journal} {Proc. Natl. Acad. Sci.
  U.S.A.}\ }\textbf {\bibinfo {volume} {108}},\ \bibinfo {pages} {20908}
  (\bibinfo {year} {2011})}\BibitemShut {NoStop}%
\bibitem [{\citenamefont {Chain}\ and\ \citenamefont
  {Arnon}(1977)}]{Chain1977}%
  \BibitemOpen
  \bibfield  {author} {\bibinfo {author} {\bibfnamefont {R.~K.}\ \bibnamefont
  {Chain}}\ and\ \bibinfo {author} {\bibfnamefont {D.~I.}\ \bibnamefont
  {Arnon}},\ }\href@noop {} {\bibfield  {journal} {\bibinfo  {journal} {Proc.
  Natl. Acad. Sci. U.S.A.}\ }\textbf {\bibinfo {volume} {74}},\ \bibinfo
  {pages} {3377} (\bibinfo {year} {1977})}\BibitemShut {NoStop}%
\bibitem [{\citenamefont {Qin}\ \emph {et~al.}(2014)\citenamefont {Qin},
  \citenamefont {Shen}, \citenamefont {Zhao},\ and\ \citenamefont
  {Yi}}]{Qin2014}%
  \BibitemOpen
  \bibfield  {author} {\bibinfo {author} {\bibfnamefont {M.}~\bibnamefont
  {Qin}}, \bibinfo {author} {\bibfnamefont {H.~Z.}\ \bibnamefont {Shen}},
  \bibinfo {author} {\bibfnamefont {X.~L.}\ \bibnamefont {Zhao}}, \ and\
  \bibinfo {author} {\bibfnamefont {X.~X.}\ \bibnamefont {Yi}},\ }\href@noop {}
  {\bibfield  {journal} {\bibinfo  {journal} {Phys. Rev. E}\ }\textbf {\bibinfo
  {volume} {90}},\ \bibinfo {pages} {042140} (\bibinfo {year}
  {2014})}\BibitemShut {NoStop}%
\bibitem [{\citenamefont {Plenio}\ and\ \citenamefont
  {Huelga}(2008)}]{Plenio2008}%
  \BibitemOpen
  \bibfield  {author} {\bibinfo {author} {\bibfnamefont {M.~B.}\ \bibnamefont
  {Plenio}}\ and\ \bibinfo {author} {\bibfnamefont {S.~F.}\ \bibnamefont
  {Huelga}},\ }\href@noop {} {\bibfield  {journal} {\bibinfo  {journal} {New J.
  Phys.}\ }\textbf {\bibinfo {volume} {10}},\ \bibinfo {pages} {113019}
  (\bibinfo {year} {2008})}\BibitemShut {NoStop}%
\bibitem [{\citenamefont {Caruso}\ \emph {et~al.}(2009)\citenamefont {Caruso},
  \citenamefont {Chin}, \citenamefont {Datta}, \citenamefont {Huelga},\ and\
  \citenamefont {Plenio}}]{Caruso2009}%
  \BibitemOpen
  \bibfield  {author} {\bibinfo {author} {\bibfnamefont {F.}~\bibnamefont
  {Caruso}}, \bibinfo {author} {\bibfnamefont {A.~W.}\ \bibnamefont {Chin}},
  \bibinfo {author} {\bibfnamefont {A.}~\bibnamefont {Datta}}, \bibinfo
  {author} {\bibfnamefont {S.~F.}\ \bibnamefont {Huelga}}, \ and\ \bibinfo
  {author} {\bibfnamefont {M.~B.}\ \bibnamefont {Plenio}},\ }\href@noop {}
  {\bibfield  {journal} {\bibinfo  {journal} {J. Chem. Phys.}\ }\textbf
  {\bibinfo {volume} {131}},\ \bibinfo {pages} {105106} (\bibinfo {year}
  {2009})}\BibitemShut {NoStop}%
\bibitem [{\citenamefont {Chin}\ \emph {et~al.}(2010)\citenamefont {Chin},
  \citenamefont {Datta}, \citenamefont {Caruso}, \citenamefont {Huelga},\ and\
  \citenamefont {Plenio}}]{Chin2010}%
  \BibitemOpen
  \bibfield  {author} {\bibinfo {author} {\bibfnamefont {A.~W.}\ \bibnamefont
  {Chin}}, \bibinfo {author} {\bibfnamefont {A.}~\bibnamefont {Datta}},
  \bibinfo {author} {\bibfnamefont {F.}~\bibnamefont {Caruso}}, \bibinfo
  {author} {\bibfnamefont {S.~F.}\ \bibnamefont {Huelga}}, \ and\ \bibinfo
  {author} {\bibfnamefont {M.~B.}\ \bibnamefont {Plenio}},\ }\href@noop {}
  {\bibfield  {journal} {\bibinfo  {journal} {New J. Phys.}\ }\textbf {\bibinfo
  {volume} {12}},\ \bibinfo {pages} {065002} (\bibinfo {year}
  {2010})}\BibitemShut {NoStop}%
\bibitem [{\citenamefont {Sinayskiy}\ \emph {et~al.}(2012)\citenamefont
  {Sinayskiy}, \citenamefont {Marais}, \citenamefont {Petruccione},\ and\
  \citenamefont {Ekert}}]{Sinayskiy2012}%
  \BibitemOpen
  \bibfield  {author} {\bibinfo {author} {\bibfnamefont {I.}~\bibnamefont
  {Sinayskiy}}, \bibinfo {author} {\bibfnamefont {A.}~\bibnamefont {Marais}},
  \bibinfo {author} {\bibfnamefont {F.}~\bibnamefont {Petruccione}}, \ and\
  \bibinfo {author} {\bibfnamefont {A.}~\bibnamefont {Ekert}},\ }\href@noop {}
  {\bibfield  {journal} {\bibinfo  {journal} {Phys. Rev. Lett.}\ }\textbf
  {\bibinfo {volume} {108}},\ \bibinfo {pages} {020602} (\bibinfo {year}
  {2012})}\BibitemShut {NoStop}%
\bibitem [{\citenamefont {Marais}\ \emph {et~al.}(2013)\citenamefont {Marais},
  \citenamefont {Sinayskiy}, \citenamefont {Kay}, \citenamefont {Petruccione},\
  and\ \citenamefont {Ekert}}]{Marais2013}%
  \BibitemOpen
  \bibfield  {author} {\bibinfo {author} {\bibfnamefont {A.}~\bibnamefont
  {Marais}}, \bibinfo {author} {\bibfnamefont {I.}~\bibnamefont {Sinayskiy}},
  \bibinfo {author} {\bibfnamefont {A.}~\bibnamefont {Kay}}, \bibinfo {author}
  {\bibfnamefont {F.}~\bibnamefont {Petruccione}}, \ and\ \bibinfo {author}
  {\bibfnamefont {A.}~\bibnamefont {Ekert}},\ }\href@noop {} {\bibfield
  {journal} {\bibinfo  {journal} {New J. Phys.}\ }\textbf {\bibinfo {volume}
  {15}},\ \bibinfo {pages} {013038} (\bibinfo {year} {2013})}\BibitemShut
  {NoStop}%
\bibitem [{\citenamefont {Wu}\ \emph {et~al.}(2010)\citenamefont {Wu},
  \citenamefont {Liu}, \citenamefont {Shen}, \citenamefont {Cao},\ and\
  \citenamefont {Silbey}}]{Wu2010}%
  \BibitemOpen
  \bibfield  {author} {\bibinfo {author} {\bibfnamefont {J.}~\bibnamefont
  {Wu}}, \bibinfo {author} {\bibfnamefont {F.}~\bibnamefont {Liu}}, \bibinfo
  {author} {\bibfnamefont {Y.}~\bibnamefont {Shen}}, \bibinfo {author}
  {\bibfnamefont {J.}~\bibnamefont {Cao}}, \ and\ \bibinfo {author}
  {\bibfnamefont {R.~J.}\ \bibnamefont {Silbey}},\ }\href@noop {} {\bibfield
  {journal} {\bibinfo  {journal} {New. J. Phys.}\ }\textbf {\bibinfo {volume}
  {12}},\ \bibinfo {pages} {105012} (\bibinfo {year} {2010})}\BibitemShut
  {NoStop}%
\bibitem [{\citenamefont {Rebentrost}\ \emph
  {et~al.}(2009{\natexlab{a}})\citenamefont {Rebentrost}, \citenamefont
  {Mohseni},\ and\ \citenamefont {Aspuru-Guzik}}]{Rebentrost2009_1}%
  \BibitemOpen
  \bibfield  {author} {\bibinfo {author} {\bibfnamefont {P.}~\bibnamefont
  {Rebentrost}}, \bibinfo {author} {\bibfnamefont {M.}~\bibnamefont {Mohseni}},
  \ and\ \bibinfo {author} {\bibfnamefont {A.}~\bibnamefont {Aspuru-Guzik}},\
  }\href@noop {} {\bibfield  {journal} {\bibinfo  {journal} {J. Phys. Chem. B}\
  }\textbf {\bibinfo {volume} {113}},\ \bibinfo {pages} {9942} (\bibinfo {year}
  {2009}{\natexlab{a}})}\BibitemShut {NoStop}%
\bibitem [{\citenamefont {Rebentrost}\ \emph
  {et~al.}(2009{\natexlab{b}})\citenamefont {Rebentrost}, \citenamefont
  {Mohseni}, \citenamefont {Kassal}, \citenamefont {Lloyd},\ and\ \citenamefont
  {Aspuru-Guzik}}]{Rebentrost2009_2}%
  \BibitemOpen
  \bibfield  {author} {\bibinfo {author} {\bibfnamefont {P.}~\bibnamefont
  {Rebentrost}}, \bibinfo {author} {\bibfnamefont {M.}~\bibnamefont {Mohseni}},
  \bibinfo {author} {\bibfnamefont {I.}~\bibnamefont {Kassal}}, \bibinfo
  {author} {\bibfnamefont {S.}~\bibnamefont {Lloyd}}, \ and\ \bibinfo {author}
  {\bibfnamefont {A.}~\bibnamefont {Aspuru-Guzik}},\ }\href@noop {} {\bibfield
  {journal} {\bibinfo  {journal} {New J. Phys.}\ }\textbf {\bibinfo {volume}
  {11}},\ \bibinfo {pages} {033003} (\bibinfo {year}
  {2009}{\natexlab{b}})}\BibitemShut {NoStop}%
\bibitem [{\citenamefont {Mohseni}\ \emph {et~al.}(2008)\citenamefont
  {Mohseni}, \citenamefont {Rebentrost}, \citenamefont {Lloyd},\ and\
  \citenamefont {Aspuru-Guzik}}]{Mohseni2008}%
  \BibitemOpen
  \bibfield  {author} {\bibinfo {author} {\bibfnamefont {M.}~\bibnamefont
  {Mohseni}}, \bibinfo {author} {\bibfnamefont {P.}~\bibnamefont {Rebentrost}},
  \bibinfo {author} {\bibfnamefont {S.}~\bibnamefont {Lloyd}}, \ and\ \bibinfo
  {author} {\bibfnamefont {A.}~\bibnamefont {Aspuru-Guzik}},\ }\href@noop {}
  {\bibfield  {journal} {\bibinfo  {journal} {J. Chem. Phys.}\ }\textbf
  {\bibinfo {volume} {129}},\ \bibinfo {pages} {174106} (\bibinfo {year}
  {2008})}\BibitemShut {NoStop}%
\bibitem [{\citenamefont {Dijkstra}\ and\ \citenamefont
  {Tanimura}(2012)}]{Dijkstra2012}%
  \BibitemOpen
  \bibfield  {author} {\bibinfo {author} {\bibfnamefont {A.~G.}\ \bibnamefont
  {Dijkstra}}\ and\ \bibinfo {author} {\bibfnamefont {Y.}~\bibnamefont
  {Tanimura}},\ }\href@noop {} {\bibfield  {journal} {\bibinfo  {journal} {New
  J. Phys.}\ }\textbf {\bibinfo {volume} {14}},\ \bibinfo {pages} {073027}
  (\bibinfo {year} {2012})}\BibitemShut {NoStop}%
\bibitem [{\citenamefont {Shabani}\ \emph {et~al.}(2012)\citenamefont
  {Shabani}, \citenamefont {Mohseni}, \citenamefont {Rabitz},\ and\
  \citenamefont {Lloyd}}]{Shabani2012}%
  \BibitemOpen
  \bibfield  {author} {\bibinfo {author} {\bibfnamefont {A.}~\bibnamefont
  {Shabani}}, \bibinfo {author} {\bibfnamefont {M.}~\bibnamefont {Mohseni}},
  \bibinfo {author} {\bibfnamefont {H.}~\bibnamefont {Rabitz}}, \ and\ \bibinfo
  {author} {\bibfnamefont {S.}~\bibnamefont {Lloyd}},\ }\href@noop {}
  {\bibfield  {journal} {\bibinfo  {journal} {Phys. Rev. E}\ }\textbf {\bibinfo
  {volume} {86}},\ \bibinfo {pages} {011915} (\bibinfo {year}
  {2012})}\BibitemShut {NoStop}%
\bibitem [{\citenamefont {Mohseni}\ \emph {et~al.}(2014)\citenamefont
  {Mohseni}, \citenamefont {Shabani}, \citenamefont {Lloyd},\ and\
  \citenamefont {Rabitz}}]{Mohseni2014}%
  \BibitemOpen
  \bibfield  {author} {\bibinfo {author} {\bibfnamefont {M.}~\bibnamefont
  {Mohseni}}, \bibinfo {author} {\bibfnamefont {A.}~\bibnamefont {Shabani}},
  \bibinfo {author} {\bibfnamefont {S.}~\bibnamefont {Lloyd}}, \ and\ \bibinfo
  {author} {\bibfnamefont {H.}~\bibnamefont {Rabitz}},\ }\href@noop {}
  {\bibfield  {journal} {\bibinfo  {journal} {J. Chem. Phys.}\ }\textbf
  {\bibinfo {volume} {140}},\ \bibinfo {pages} {035102} (\bibinfo {year}
  {2014})}\BibitemShut {NoStop}%
\bibitem [{\citenamefont {Tanimura}\ and\ \citenamefont
  {Kubo}(1989{\natexlab{a}})}]{Tanimura1989_1}%
  \BibitemOpen
  \bibfield  {author} {\bibinfo {author} {\bibfnamefont {Y.}~\bibnamefont
  {Tanimura}}\ and\ \bibinfo {author} {\bibfnamefont {R.}~\bibnamefont
  {Kubo}},\ }\href@noop {} {\bibfield  {journal} {\bibinfo  {journal} {J. Phys.
  Soc. Jpn.}\ }\textbf {\bibinfo {volume} {58}},\ \bibinfo {pages} {101}
  (\bibinfo {year} {1989}{\natexlab{a}})}\BibitemShut {NoStop}%
\bibitem [{\citenamefont {Tanimura}\ and\ \citenamefont
  {Kubo}(1989{\natexlab{b}})}]{Tanimura1989_2}%
  \BibitemOpen
  \bibfield  {author} {\bibinfo {author} {\bibfnamefont {Y.}~\bibnamefont
  {Tanimura}}\ and\ \bibinfo {author} {\bibfnamefont {R.}~\bibnamefont
  {Kubo}},\ }\href@noop {} {\bibfield  {journal} {\bibinfo  {journal} {J. Phys.
  Soc. Jpn.}\ }\textbf {\bibinfo {volume} {58}},\ \bibinfo {pages} {1199}
  (\bibinfo {year} {1989}{\natexlab{b}})}\BibitemShut {NoStop}%
\bibitem [{\citenamefont {Tanimura}(1990)}]{Tanimura1990}%
  \BibitemOpen
  \bibfield  {author} {\bibinfo {author} {\bibfnamefont {Y.}~\bibnamefont
  {Tanimura}},\ }\href@noop {} {\bibfield  {journal} {\bibinfo  {journal}
  {Phys. Rev. A}\ }\textbf {\bibinfo {volume} {41}},\ \bibinfo {pages} {6676}
  (\bibinfo {year} {1990})}\BibitemShut {NoStop}%
\bibitem [{\citenamefont {Ishizaki}\ and\ \citenamefont
  {Fleming}(2009{\natexlab{a}})}]{Ishizaki2009_2}%
  \BibitemOpen
  \bibfield  {author} {\bibinfo {author} {\bibfnamefont {A.}~\bibnamefont
  {Ishizaki}}\ and\ \bibinfo {author} {\bibfnamefont {G.~R.}\ \bibnamefont
  {Fleming}},\ }\href@noop {} {\bibfield  {journal} {\bibinfo  {journal} {J.
  Chem. Phys.}\ }\textbf {\bibinfo {volume} {130}},\ \bibinfo {pages} {234111}
  (\bibinfo {year} {2009}{\natexlab{a}})}\BibitemShut {NoStop}%
\bibitem [{\citenamefont {Ishizaki}\ and\ \citenamefont
  {Fleming}(2009{\natexlab{b}})}]{Ishizaki2009_1}%
  \BibitemOpen
  \bibfield  {author} {\bibinfo {author} {\bibfnamefont {A.}~\bibnamefont
  {Ishizaki}}\ and\ \bibinfo {author} {\bibfnamefont {G.~R.}\ \bibnamefont
  {Fleming}},\ }\href@noop {} {\bibfield  {journal} {\bibinfo  {journal} {Proc.
  Natl. Acad. Sci. U.S.A.}\ }\textbf {\bibinfo {volume} {106}},\ \bibinfo
  {pages} {17255} (\bibinfo {year} {2009}{\natexlab{b}})}\BibitemShut {NoStop}%
\bibitem [{\citenamefont {Paz}\ and\ \citenamefont
  {Roncaglia}(2008)}]{Paz2008}%
  \BibitemOpen
  \bibfield  {author} {\bibinfo {author} {\bibfnamefont {J.~P.}\ \bibnamefont
  {Paz}}\ and\ \bibinfo {author} {\bibfnamefont {A.~J.}\ \bibnamefont
  {Roncaglia}},\ }\href@noop {} {\bibfield  {journal} {\bibinfo  {journal}
  {Phys. Rev. Lett.}\ }\textbf {\bibinfo {volume} {100}},\ \bibinfo {pages}
  {220401} (\bibinfo {year} {2008})}\BibitemShut {NoStop}%
\bibitem [{\citenamefont {Valido}\ \emph
  {et~al.}(2013{\natexlab{a}})\citenamefont {Valido}, \citenamefont {Correa},\
  and\ \citenamefont {Alonso}}]{Valido2013_1}%
  \BibitemOpen
  \bibfield  {author} {\bibinfo {author} {\bibfnamefont {A.~A.}\ \bibnamefont
  {Valido}}, \bibinfo {author} {\bibfnamefont {L.~A.}\ \bibnamefont {Correa}},
  \ and\ \bibinfo {author} {\bibfnamefont {D.}~\bibnamefont {Alonso}},\
  }\href@noop {} {\bibfield  {journal} {\bibinfo  {journal} {Phys. Rev. A}\
  }\textbf {\bibinfo {volume} {88}},\ \bibinfo {pages} {012309} (\bibinfo
  {year} {2013}{\natexlab{a}})}\BibitemShut {NoStop}%
\bibitem [{\citenamefont {Valido}\ \emph
  {et~al.}(2013{\natexlab{b}})\citenamefont {Valido}, \citenamefont {Alonso},\
  and\ \citenamefont {Kohler}}]{Valido2013_2}%
  \BibitemOpen
  \bibfield  {author} {\bibinfo {author} {\bibfnamefont {A.~A.}\ \bibnamefont
  {Valido}}, \bibinfo {author} {\bibfnamefont {D.}~\bibnamefont {Alonso}}, \
  and\ \bibinfo {author} {\bibfnamefont {S.}~\bibnamefont {Kohler}},\
  }\href@noop {} {\bibfield  {journal} {\bibinfo  {journal} {Phys. Rev. A}\
  }\textbf {\bibinfo {volume} {88}},\ \bibinfo {pages} {042303} (\bibinfo
  {year} {2013}{\natexlab{b}})}\BibitemShut {NoStop}%
\bibitem [{\citenamefont {Bellomo}\ \emph {et~al.}(2007)\citenamefont
  {Bellomo}, \citenamefont {Lo~Franco},\ and\ \citenamefont
  {Compagno}}]{Bellomo2007}%
  \BibitemOpen
  \bibfield  {author} {\bibinfo {author} {\bibfnamefont {B.}~\bibnamefont
  {Bellomo}}, \bibinfo {author} {\bibfnamefont {R.}~\bibnamefont {Lo~Franco}},
  \ and\ \bibinfo {author} {\bibfnamefont {G.}~\bibnamefont {Compagno}},\
  }\href@noop {} {\bibfield  {journal} {\bibinfo  {journal} {Phys. Rev. Lett.}\
  }\textbf {\bibinfo {volume} {99}},\ \bibinfo {pages} {160502} (\bibinfo
  {year} {2007})}\BibitemShut {NoStop}%
\bibitem [{\citenamefont {Huelga}\ \emph {et~al.}(2012)\citenamefont {Huelga},
  \citenamefont {Rivas},\ and\ \citenamefont {Plenio}}]{Huelga2012}%
  \BibitemOpen
  \bibfield  {author} {\bibinfo {author} {\bibfnamefont {S.~F.}\ \bibnamefont
  {Huelga}}, \bibinfo {author} {\bibfnamefont {{\'A}.}~\bibnamefont {Rivas}}, \
  and\ \bibinfo {author} {\bibfnamefont {M.~B.}\ \bibnamefont {Plenio}},\
  }\href@noop {} {\bibfield  {journal} {\bibinfo  {journal} {Phys. Rev. Lett.}\
  }\textbf {\bibinfo {volume} {108}},\ \bibinfo {pages} {160402} (\bibinfo
  {year} {2012})}\BibitemShut {NoStop}%
\bibitem [{\citenamefont {Maniscalco}\ \emph {et~al.}(2007)\citenamefont
  {Maniscalco}, \citenamefont {Olivares},\ and\ \citenamefont
  {Paris}}]{Maniscalco2007}%
  \BibitemOpen
  \bibfield  {author} {\bibinfo {author} {\bibfnamefont {S.}~\bibnamefont
  {Maniscalco}}, \bibinfo {author} {\bibfnamefont {S.}~\bibnamefont
  {Olivares}}, \ and\ \bibinfo {author} {\bibfnamefont {M.~G.~A.}\ \bibnamefont
  {Paris}},\ }\href@noop {} {\bibfield  {journal} {\bibinfo  {journal} {Phys.
  Rev. A}\ }\textbf {\bibinfo {volume} {75}},\ \bibinfo {pages} {062119}
  (\bibinfo {year} {2007})}\BibitemShut {NoStop}%
\bibitem [{\citenamefont {Bylicka}\ \emph {et~al.}(2014)\citenamefont
  {Bylicka}, \citenamefont {Chru{\'s}ci{\'n}ski},\ and\ \citenamefont
  {Maniscalco}}]{Bylicka2014}%
  \BibitemOpen
  \bibfield  {author} {\bibinfo {author} {\bibfnamefont {B.}~\bibnamefont
  {Bylicka}}, \bibinfo {author} {\bibfnamefont {D.}~\bibnamefont
  {Chru{\'s}ci{\'n}ski}}, \ and\ \bibinfo {author} {\bibfnamefont
  {S.}~\bibnamefont {Maniscalco}},\ }\href@noop {} {\bibfield  {journal}
  {\bibinfo  {journal} {Sci. Rep.}\ }\textbf {\bibinfo {volume} {4}},\ \bibinfo
  {pages} {5720} (\bibinfo {year} {2014})}\BibitemShut {NoStop}%
\bibitem [{\citenamefont {Caruso}\ \emph {et~al.}(2014)\citenamefont {Caruso},
  \citenamefont {Giovannetti}, \citenamefont {Lupo},\ and\ \citenamefont
  {Mancini}}]{Caruso2014}%
  \BibitemOpen
  \bibfield  {author} {\bibinfo {author} {\bibfnamefont {F.}~\bibnamefont
  {Caruso}}, \bibinfo {author} {\bibfnamefont {V.}~\bibnamefont {Giovannetti}},
  \bibinfo {author} {\bibfnamefont {C.}~\bibnamefont {Lupo}}, \ and\ \bibinfo
  {author} {\bibfnamefont {S.}~\bibnamefont {Mancini}},\ }\href@noop {}
  {\bibfield  {journal} {\bibinfo  {journal} {Rev. Mod. Phys.}\ }\textbf
  {\bibinfo {volume} {86}},\ \bibinfo {pages} {1203} (\bibinfo {year}
  {2014})}\BibitemShut {NoStop}%
\bibitem [{\citenamefont {Laine}\ \emph {et~al.}(2014)\citenamefont {Laine},
  \citenamefont {Breuer},\ and\ \citenamefont {Piilo}}]{Laine2014}%
  \BibitemOpen
  \bibfield  {author} {\bibinfo {author} {\bibfnamefont {E.-M.}\ \bibnamefont
  {Laine}}, \bibinfo {author} {\bibfnamefont {H.-P.}\ \bibnamefont {Breuer}}, \
  and\ \bibinfo {author} {\bibfnamefont {J.}~\bibnamefont {Piilo}},\
  }\href@noop {} {\bibfield  {journal} {\bibinfo  {journal} {Sci. Rep.}\
  }\textbf {\bibinfo {volume} {4}},\ \bibinfo {pages} {4620} (\bibinfo {year}
  {2014})}\BibitemShut {NoStop}%
\bibitem [{\citenamefont {Liu}\ \emph {et~al.}(2013)\citenamefont {Liu},
  \citenamefont {Cao}, \citenamefont {Huang}, \citenamefont {Li}, \citenamefont
  {Guo}, \citenamefont {Laine}, \citenamefont {Breuer},\ and\ \citenamefont
  {Piilo}}]{Liu2013}%
  \BibitemOpen
  \bibfield  {author} {\bibinfo {author} {\bibfnamefont {B.-H.}\ \bibnamefont
  {Liu}}, \bibinfo {author} {\bibfnamefont {D.-Y.}\ \bibnamefont {Cao}},
  \bibinfo {author} {\bibfnamefont {Y.-F.}\ \bibnamefont {Huang}}, \bibinfo
  {author} {\bibfnamefont {C.-F.}\ \bibnamefont {Li}}, \bibinfo {author}
  {\bibfnamefont {G.-C.}\ \bibnamefont {Guo}}, \bibinfo {author} {\bibfnamefont
  {E.-M.}\ \bibnamefont {Laine}}, \bibinfo {author} {\bibfnamefont {H.-P.}\
  \bibnamefont {Breuer}}, \ and\ \bibinfo {author} {\bibfnamefont
  {J.}~\bibnamefont {Piilo}},\ }\href@noop {} {\bibfield  {journal} {\bibinfo
  {journal} {Sci. Rep.}\ }\textbf {\bibinfo {volume} {3}},\ \bibinfo {pages}
  {1781} (\bibinfo {year} {2013})}\BibitemShut {NoStop}%
\bibitem [{\citenamefont {Thorwart}\ \emph {et~al.}(2009)\citenamefont
  {Thorwart}, \citenamefont {Eckel}, \citenamefont {Reina}, \citenamefont
  {Nalbach},\ and\ \citenamefont {Weiss}}]{Thorwart2009}%
  \BibitemOpen
  \bibfield  {author} {\bibinfo {author} {\bibfnamefont {M.}~\bibnamefont
  {Thorwart}}, \bibinfo {author} {\bibfnamefont {J.}~\bibnamefont {Eckel}},
  \bibinfo {author} {\bibfnamefont {J.~H.}\ \bibnamefont {Reina}}, \bibinfo
  {author} {\bibfnamefont {P.}~\bibnamefont {Nalbach}}, \ and\ \bibinfo
  {author} {\bibfnamefont {S.}~\bibnamefont {Weiss}},\ }\href@noop {}
  {\bibfield  {journal} {\bibinfo  {journal} {Chem. Phys. Lett.}\ }\textbf
  {\bibinfo {volume} {478}},\ \bibinfo {pages} {234} (\bibinfo {year}
  {2009})}\BibitemShut {NoStop}%
\bibitem [{\citenamefont {Chen}\ \emph {et~al.}(2014)\citenamefont {Chen},
  \citenamefont {Lien}, \citenamefont {Chi-Chuan},\ and\ \citenamefont
  {Chen}}]{Chen2014}%
  \BibitemOpen
  \bibfield  {author} {\bibinfo {author} {\bibfnamefont {H.-B.}\ \bibnamefont
  {Chen}}, \bibinfo {author} {\bibfnamefont {J.-Y.}\ \bibnamefont {Lien}},
  \bibinfo {author} {\bibfnamefont {H.}~\bibnamefont {Chi-Chuan}}, \ and\
  \bibinfo {author} {\bibfnamefont {Y.-N.}\ \bibnamefont {Chen}},\ }\href@noop
  {} {\bibfield  {journal} {\bibinfo  {journal} {Phys. Rev. E}\ }\textbf
  {\bibinfo {volume} {89}},\ \bibinfo {pages} {042147} (\bibinfo {year}
  {2014})}\BibitemShut {NoStop}%
\bibitem [{\citenamefont {Rebentrost}\ and\ \citenamefont
  {Aspuru-Guzik}(2011)}]{Rebentrost2011}%
  \BibitemOpen
  \bibfield  {author} {\bibinfo {author} {\bibfnamefont {P.}~\bibnamefont
  {Rebentrost}}\ and\ \bibinfo {author} {\bibfnamefont {A.}~\bibnamefont
  {Aspuru-Guzik}},\ }\href@noop {} {\bibfield  {journal} {\bibinfo  {journal}
  {J. Chem. Phys.}\ }\textbf {\bibinfo {volume} {134}},\ \bibinfo {pages}
  {101103} (\bibinfo {year} {2011})}\BibitemShut {NoStop}%
\bibitem [{\citenamefont {Mujica-Martinez}\ \emph {et~al.}(2013)\citenamefont
  {Mujica-Martinez}, \citenamefont {Nalbach},\ and\ \citenamefont
  {Thorwart}}]{Mujica-Martinez2013}%
  \BibitemOpen
  \bibfield  {author} {\bibinfo {author} {\bibfnamefont {C.~A.}\ \bibnamefont
  {Mujica-Martinez}}, \bibinfo {author} {\bibfnamefont {P.}~\bibnamefont
  {Nalbach}}, \ and\ \bibinfo {author} {\bibfnamefont {M.}~\bibnamefont
  {Thorwart}},\ }\href@noop {} {\bibfield  {journal} {\bibinfo  {journal}
  {Phys. Rev. E}\ }\textbf {\bibinfo {volume} {88}},\ \bibinfo {pages} {062719}
  (\bibinfo {year} {2013})}\BibitemShut {NoStop}%
\bibitem [{\citenamefont {Breuer}\ \emph {et~al.}(2009)\citenamefont {Breuer},
  \citenamefont {Laine},\ and\ \citenamefont {Piilo}}]{Breuer2009}%
  \BibitemOpen
  \bibfield  {author} {\bibinfo {author} {\bibfnamefont {H.-P.}\ \bibnamefont
  {Breuer}}, \bibinfo {author} {\bibfnamefont {E.-M.}\ \bibnamefont {Laine}}, \
  and\ \bibinfo {author} {\bibfnamefont {J.}~\bibnamefont {Piilo}},\
  }\href@noop {} {\bibfield  {journal} {\bibinfo  {journal} {Phys. Rev. Lett.}\
  }\textbf {\bibinfo {volume} {103}},\ \bibinfo {pages} {210401} (\bibinfo
  {year} {2009})}\BibitemShut {NoStop}%
\bibitem [{\citenamefont {Rivas}\ \emph {et~al.}(2010)\citenamefont {Rivas},
  \citenamefont {Huelga},\ and\ \citenamefont {Plenio}}]{Rivas2010}%
  \BibitemOpen
  \bibfield  {author} {\bibinfo {author} {\bibfnamefont {{\'A}.}~\bibnamefont
  {Rivas}}, \bibinfo {author} {\bibfnamefont {S.~F.}\ \bibnamefont {Huelga}}, \
  and\ \bibinfo {author} {\bibfnamefont {M.~B.}\ \bibnamefont {Plenio}},\
  }\href@noop {} {\bibfield  {journal} {\bibinfo  {journal} {Phys. Rev. Lett.}\
  }\textbf {\bibinfo {volume} {105}},\ \bibinfo {pages} {050403} (\bibinfo
  {year} {2010})}\BibitemShut {NoStop}%
\bibitem [{\citenamefont {Rajogopal}\ \emph {et~al.}(2010)\citenamefont
  {Rajogopal}, \citenamefont {Usha~Devi},\ and\ \citenamefont
  {Rendell}}]{Rajogopal2010}%
  \BibitemOpen
  \bibfield  {author} {\bibinfo {author} {\bibfnamefont {A.~K.}\ \bibnamefont
  {Rajogopal}}, \bibinfo {author} {\bibfnamefont {A.~R.}\ \bibnamefont
  {Usha~Devi}}, \ and\ \bibinfo {author} {\bibfnamefont {R.~W.}\ \bibnamefont
  {Rendell}},\ }\href@noop {} {\bibfield  {journal} {\bibinfo  {journal} {Phys.
  Rev. A}\ }\textbf {\bibinfo {volume} {82}},\ \bibinfo {pages} {042107}
  (\bibinfo {year} {2010})}\BibitemShut {NoStop}%
\bibitem [{\citenamefont {Baumgratz}\ \emph {et~al.}(2014)\citenamefont
  {Baumgratz}, \citenamefont {Cramer},\ and\ \citenamefont
  {Plenio}}]{Baumgratz2014}%
  \BibitemOpen
  \bibfield  {author} {\bibinfo {author} {\bibfnamefont {T.}~\bibnamefont
  {Baumgratz}}, \bibinfo {author} {\bibfnamefont {M.}~\bibnamefont {Cramer}}, \
  and\ \bibinfo {author} {\bibfnamefont {M.~B.}\ \bibnamefont {Plenio}},\
  }\href@noop {} {\bibfield  {journal} {\bibinfo  {journal} {Phys. Rev. Lett.}\
  }\textbf {\bibinfo {volume} {113}},\ \bibinfo {pages} {140401} (\bibinfo
  {year} {2014})}\BibitemShut {NoStop}%
\bibitem [{\citenamefont {Renger}\ \emph {et~al.}(2001)\citenamefont {Renger},
  \citenamefont {May},\ and\ \citenamefont {K{\"u}hn}}]{Renger2001}%
  \BibitemOpen
  \bibfield  {author} {\bibinfo {author} {\bibfnamefont {T.}~\bibnamefont
  {Renger}}, \bibinfo {author} {\bibfnamefont {V.}~\bibnamefont {May}}, \ and\
  \bibinfo {author} {\bibfnamefont {O.}~\bibnamefont {K{\"u}hn}},\ }\href@noop
  {} {\bibfield  {journal} {\bibinfo  {journal} {Phys. Rep.}\ }\textbf
  {\bibinfo {volume} {343}},\ \bibinfo {pages} {137} (\bibinfo {year}
  {2001})}\BibitemShut {NoStop}%
\bibitem [{\citenamefont {Fenna}\ \emph {et~al.}(1974)\citenamefont {Fenna},
  \citenamefont {Matthews}, \citenamefont {Olson},\ and\ \citenamefont
  {Shaw}}]{Fenna1974}%
  \BibitemOpen
  \bibfield  {author} {\bibinfo {author} {\bibfnamefont {R.~E.}\ \bibnamefont
  {Fenna}}, \bibinfo {author} {\bibfnamefont {B.~W.}\ \bibnamefont {Matthews}},
  \bibinfo {author} {\bibfnamefont {J.~M.}\ \bibnamefont {Olson}}, \ and\
  \bibinfo {author} {\bibfnamefont {E.~K.}\ \bibnamefont {Shaw}},\ }\href@noop
  {} {\bibfield  {journal} {\bibinfo  {journal} {J. Mol. Biol.}\ }\textbf
  {\bibinfo {volume} {84}},\ \bibinfo {pages} {231} (\bibinfo {year}
  {1974})}\BibitemShut {NoStop}%
\bibitem [{\citenamefont {Caruso}\ \emph {et~al.}(2010)\citenamefont {Caruso},
  \citenamefont {Chin}, \citenamefont {Datta}, \citenamefont {Huelga},\ and\
  \citenamefont {Plenio}}]{Caruso2010}%
  \BibitemOpen
  \bibfield  {author} {\bibinfo {author} {\bibfnamefont {S.}~\bibnamefont
  {Caruso}}, \bibinfo {author} {\bibfnamefont {A.~W.}\ \bibnamefont {Chin}},
  \bibinfo {author} {\bibfnamefont {A.}~\bibnamefont {Datta}}, \bibinfo
  {author} {\bibfnamefont {S.~F.}\ \bibnamefont {Huelga}}, \ and\ \bibinfo
  {author} {\bibfnamefont {M.~B.}\ \bibnamefont {Plenio}},\ }\href@noop {}
  {\bibfield  {journal} {\bibinfo  {journal} {Phys. Rev. A}\ }\textbf {\bibinfo
  {volume} {81}},\ \bibinfo {pages} {062346} (\bibinfo {year}
  {2010})}\BibitemShut {NoStop}%
\bibitem [{\citenamefont {Sarovar}\ \emph {et~al.}(2010)\citenamefont
  {Sarovar}, \citenamefont {Ishizaki}, \citenamefont {Fleming},\ and\
  \citenamefont {Whaley}}]{Sarovar2010}%
  \BibitemOpen
  \bibfield  {author} {\bibinfo {author} {\bibfnamefont {M.}~\bibnamefont
  {Sarovar}}, \bibinfo {author} {\bibfnamefont {A.}~\bibnamefont {Ishizaki}},
  \bibinfo {author} {\bibfnamefont {G.~R.}\ \bibnamefont {Fleming}}, \ and\
  \bibinfo {author} {\bibfnamefont {K.~B.}\ \bibnamefont {Whaley}},\
  }\href@noop {} {\bibfield  {journal} {\bibinfo  {journal} {Nat. Phys.}\
  }\textbf {\bibinfo {volume} {6}},\ \bibinfo {pages} {462} (\bibinfo {year}
  {2010})}\BibitemShut {NoStop}%
\bibitem [{\citenamefont {Fassioli}\ and\ \citenamefont
  {Olaya-Castro}(2010)}]{Fassioli2010}%
  \BibitemOpen
  \bibfield  {author} {\bibinfo {author} {\bibfnamefont {F.}~\bibnamefont
  {Fassioli}}\ and\ \bibinfo {author} {\bibfnamefont {A.}~\bibnamefont
  {Olaya-Castro}},\ }\href@noop {} {\bibfield  {journal} {\bibinfo  {journal}
  {New J. Phys.}\ }\textbf {\bibinfo {volume} {12}},\ \bibinfo {pages} {085006}
  (\bibinfo {year} {2010})}\BibitemShut {NoStop}%
\bibitem [{\citenamefont {Baker}\ and\ \citenamefont
  {Habershon}(2015)}]{Baker2015}%
  \BibitemOpen
  \bibfield  {author} {\bibinfo {author} {\bibfnamefont {L.~A.}\ \bibnamefont
  {Baker}}\ and\ \bibinfo {author} {\bibfnamefont {S.}~\bibnamefont
  {Habershon}},\ }\href@noop {} {\bibfield  {journal} {\bibinfo  {journal} {J.
  Chem. Phys.}\ }\textbf {\bibinfo {volume} {143}},\ \bibinfo {pages} {105101}
  (\bibinfo {year} {2015})}\BibitemShut {NoStop}%
\bibitem [{\citenamefont {Chen}\ \emph {et~al.}(2013)\citenamefont {Chen},
  \citenamefont {Lambert}, \citenamefont {Li}, \citenamefont {Chen},\ and\
  \citenamefont {Nori}}]{Chen2013}%
  \BibitemOpen
  \bibfield  {author} {\bibinfo {author} {\bibfnamefont {G.-Y.}\ \bibnamefont
  {Chen}}, \bibinfo {author} {\bibfnamefont {N.}~\bibnamefont {Lambert}},
  \bibinfo {author} {\bibfnamefont {C.-M.}\ \bibnamefont {Li}}, \bibinfo
  {author} {\bibfnamefont {Y.-N.}\ \bibnamefont {Chen}}, \ and\ \bibinfo
  {author} {\bibfnamefont {F.}~\bibnamefont {Nori}},\ }\href@noop {} {\bibfield
   {journal} {\bibinfo  {journal} {Phys. Rev. E}\ }\textbf {\bibinfo {volume}
  {88}},\ \bibinfo {pages} {032120} (\bibinfo {year} {2013})}\BibitemShut
  {NoStop}%
\bibitem [{\citenamefont {Shabani}\ \emph {et~al.}(2014)\citenamefont
  {Shabani}, \citenamefont {Mohseni}, \citenamefont {Rabitz},\ and\
  \citenamefont {Lloyd}}]{Shabani2014}%
  \BibitemOpen
  \bibfield  {author} {\bibinfo {author} {\bibfnamefont {A.}~\bibnamefont
  {Shabani}}, \bibinfo {author} {\bibfnamefont {M.}~\bibnamefont {Mohseni}},
  \bibinfo {author} {\bibfnamefont {H.}~\bibnamefont {Rabitz}}, \ and\ \bibinfo
  {author} {\bibfnamefont {S.}~\bibnamefont {Lloyd}},\ }\href@noop {}
  {\bibfield  {journal} {\bibinfo  {journal} {Phys. Rev. E}\ }\textbf {\bibinfo
  {volume} {89}},\ \bibinfo {pages} {042706} (\bibinfo {year}
  {2014})}\BibitemShut {NoStop}%
\bibitem [{\citenamefont {Wu}\ \emph {et~al.}(2012)\citenamefont {Wu},
  \citenamefont {Liu}, \citenamefont {Ma}, \citenamefont {Silbey},\ and\
  \citenamefont {Cao}}]{Wu2012}%
  \BibitemOpen
  \bibfield  {author} {\bibinfo {author} {\bibfnamefont {J.}~\bibnamefont
  {Wu}}, \bibinfo {author} {\bibfnamefont {F.}~\bibnamefont {Liu}}, \bibinfo
  {author} {\bibfnamefont {J.}~\bibnamefont {Ma}}, \bibinfo {author}
  {\bibfnamefont {R.~J.}\ \bibnamefont {Silbey}}, \ and\ \bibinfo {author}
  {\bibfnamefont {J.}~\bibnamefont {Cao}},\ }\href@noop {} {\bibfield
  {journal} {\bibinfo  {journal} {J. Chem. Phys.}\ }\textbf {\bibinfo {volume}
  {137}},\ \bibinfo {pages} {174111} (\bibinfo {year} {2012})}\BibitemShut
  {NoStop}%
\bibitem [{\citenamefont {Bengtson}\ and\ \citenamefont
  {Sj{\"o}qvist}(2017)}]{Bengtson2017}%
  \BibitemOpen
  \bibfield  {author} {\bibinfo {author} {\bibfnamefont {C.}~\bibnamefont
  {Bengtson}}\ and\ \bibinfo {author} {\bibfnamefont {E.}~\bibnamefont
  {Sj{\"o}qvist}},\ }\href@noop {} {\bibfield  {journal} {\bibinfo  {journal}
  {New J. Phys.}\ }\textbf {\bibinfo {volume} {19}},\ \bibinfo {pages} {113015}
  (\bibinfo {year} {2017})}\BibitemShut {NoStop}%
\bibitem [{\citenamefont {Nielsen}\ and\ \citenamefont
  {Chuang}(2010)}]{Nielsen}%
  \BibitemOpen
  \bibfield  {author} {\bibinfo {author} {\bibfnamefont {M.~A.}\ \bibnamefont
  {Nielsen}}\ and\ \bibinfo {author} {\bibfnamefont {I.~L.}\ \bibnamefont
  {Chuang}},\ }\href@noop {} {\emph {\bibinfo {title} {Quantum Computation and
  Quantum Information}}}\ (\bibinfo  {publisher} {Cambridge University Press,
  Cambridge, England},\ \bibinfo {year} {2010})\BibitemShut {NoStop}%
\bibitem [{\citenamefont {Ruskai}(1994)}]{Ruskai1994}%
  \BibitemOpen
  \bibfield  {author} {\bibinfo {author} {\bibfnamefont {M.~B.}\ \bibnamefont
  {Ruskai}},\ }\href@noop {} {\bibfield  {journal} {\bibinfo  {journal} {Rev.
  Math. Phys.}\ }\textbf {\bibinfo {volume} {6}},\ \bibinfo {pages} {1147}
  (\bibinfo {year} {1994})}\BibitemShut {NoStop}%
\bibitem [{\citenamefont {Wi{\ss}mann}\ \emph {et~al.}(2012)\citenamefont
  {Wi{\ss}mann}, \citenamefont {Karlsson}, \citenamefont {Laine}, \citenamefont
  {Piilo},\ and\ \citenamefont {Breuer}}]{Wissmann2012}%
  \BibitemOpen
  \bibfield  {author} {\bibinfo {author} {\bibfnamefont {S.}~\bibnamefont
  {Wi{\ss}mann}}, \bibinfo {author} {\bibfnamefont {A.}~\bibnamefont
  {Karlsson}}, \bibinfo {author} {\bibfnamefont {E.-M.}\ \bibnamefont {Laine}},
  \bibinfo {author} {\bibfnamefont {J.}~\bibnamefont {Piilo}}, \ and\ \bibinfo
  {author} {\bibfnamefont {H.-P.}\ \bibnamefont {Breuer}},\ }\href@noop {}
  {\bibfield  {journal} {\bibinfo  {journal} {Phys. Rev. A}\ }\textbf {\bibinfo
  {volume} {86}},\ \bibinfo {pages} {062108} (\bibinfo {year}
  {2012})}\BibitemShut {NoStop}%
\bibitem [{\citenamefont {Bronzan}(1988)}]{Bronzan1988}%
  \BibitemOpen
  \bibfield  {author} {\bibinfo {author} {\bibfnamefont {J.~B.}\ \bibnamefont
  {Bronzan}},\ }\href@noop {} {\bibfield  {journal} {\bibinfo  {journal} {Phys.
  Rev. D}\ }\textbf {\bibinfo {volume} {38}},\ \bibinfo {pages} {1994}
  (\bibinfo {year} {1988})}\BibitemShut {NoStop}%
\bibitem [{\citenamefont {Wootters}(1998)}]{Wootters1998}%
  \BibitemOpen
  \bibfield  {author} {\bibinfo {author} {\bibfnamefont {W.~K.}\ \bibnamefont
  {Wootters}},\ }\href@noop {} {\bibfield  {journal} {\bibinfo  {journal}
  {Phys. Rev. Lett.}\ }\textbf {\bibinfo {volume} {80}},\ \bibinfo {pages}
  {2245} (\bibinfo {year} {1998})}\BibitemShut {NoStop}%
\bibitem [{\citenamefont {Suba{\c s}i}\ \emph {et~al.}(2012)\citenamefont
  {Suba{\c s}i}, \citenamefont {Fleming}, \citenamefont {Taylor},\ and\
  \citenamefont {Hu}}]{Subasi2012}%
  \BibitemOpen
  \bibfield  {author} {\bibinfo {author} {\bibfnamefont {Y.}~\bibnamefont
  {Suba{\c s}i}}, \bibinfo {author} {\bibfnamefont {C.~H.}\ \bibnamefont
  {Fleming}}, \bibinfo {author} {\bibfnamefont {J.~M.}\ \bibnamefont {Taylor}},
  \ and\ \bibinfo {author} {\bibfnamefont {B.~L.}\ \bibnamefont {Hu}},\
  }\href@noop {} {\bibfield  {journal} {\bibinfo  {journal} {Phys. Rev. E}\
  }\textbf {\bibinfo {volume} {86}},\ \bibinfo {pages} {061132} (\bibinfo
  {year} {2012})}\BibitemShut {NoStop}%
\end{thebibliography}%

\end{document}